\pdfoutput=1

\documentclass[11pt,nofootinbib]{article}
\usepackage{amsmath,amssymb,mathtools}
\usepackage{graphicx}
\usepackage[colorlinks=true,linkcolor=blue,citecolor=blue,urlcolor=blue,linktocpage=true]{hyperref}
\usepackage{subfigure}
\usepackage{comment}
\usepackage{tensor}
\usepackage{cases}
\usepackage[utf8]{inputenc}
\usepackage{cite}
\usepackage[font=small,labelfont=bf]{caption}

\numberwithin{equation}{section}

\begin{document}

\providecommand{\abs}[1]{\lvert#1\rvert}
\providecommand{\bd}[1]{\boldsymbol{#1}}
\providecommand{\ro}[1]{\mathrm{#1}}

\begin{titlepage}

\setcounter{page}{1} \baselineskip=15.5pt \thispagestyle{empty}

\begin{flushright}
\end{flushright}
\vfil

\bigskip
\begin{center}
 {\LARGE \textbf{Monopole-Antimonopole Pair Production}}\\
 \medskip
 {\LARGE \textbf{in Primordial Magnetic Fields}}
\vskip 15pt
\end{center}

\vspace{0.5cm}
\begin{center}
{\large
Takeshi Kobayashi
}\end{center}

\vspace{0.3cm}

\begin{center}
\textit{Kobayashi-Maskawa Institute for the Origin of Particles and the Universe,\\ Nagoya University, Nagoya 464-8602, Japan}\\

\vskip 14pt
E-mail:
 \texttt{\href{mailto:takeshi@kmi.nagoya-u.ac.jp}{takeshi@kmi.nagoya-u.ac.jp}}
\end{center} 



\vspace{1cm}

\noindent
We show that monopoles can be pair produced by cosmological magnetic fields in the early universe. The pair production gives rise to relic monopoles, and at the same time induces a self-screening of the magnetic fields. By studying these effects we derive limits on the monopole mass, and also on the initial amplitude of primordial magnetic fields. Monopoles of GUT scale mass can even be produced if primordial magnetic fields exist at sufficiently high redshifts.
\vfil

\end{titlepage}

\newpage
\setcounter{tocdepth}{2}
\tableofcontents

\section{Introduction}
\label{sec:intro}

The hypothesis that magnetic monopoles exist, albeit without any
experimental evidence, has long been a subject of intense research.
Monopoles would symmetrize Maxwell's equations, and
moreover their existence would be tied to 
the observed quantization of electric charge through 
the Dirac quantization condition 
$e g= 2 \pi n $, $n \in \mathbb{Z}$~\cite{Dirac:1931kp}.
Besides the possibility that monopoles are elementary particles, 
they can be realized as topological solitons in spontaneously
broken gauge theories as shown by 't Hooft and
Polyakov~\cite{tHooft:1974kcl,Polyakov:1974ek}.
The fact that such soliton solutions are contained in any Grand Unified
gauge Theory (GUT) in which the electromagnetic U(1) is embedded in a
semi-simple gauge group, makes monopoles
an inevitable prediction of grand unification.

Despite the strong theoretical support for their existence, 
monopoles remain elusive in experimental searches.
One reason is that their masses are expected to be superheavy.
Solitonic monopoles have masses of the order of the symmetry breaking
scale, which for GUT monopoles is typically $10^{16}\, \ro{GeV}$ and
thus is far beyond the reach of terrestrial colliders.  
This does depend on the model, and much lighter monopoles can
also arise, for instance, in theories with several stages of symmetry
breaking~\cite{Lazarides:1980va}.
However even with a small mass, producing solitonic monopoles at
colliders has been argued to be strongly suppressed due to their high
degree of compositeness~\cite{Witten:1979kh,Drukier:1981fq}. 
It is also important to note that the computation of the production
cross section of monopoles is a challenging task in itself.
This is because the Dirac quantization condition demands $\abs{g} \gg 1$
for $\abs{e} \ll 1$, rendering monopoles strongly coupled and
perturbation theory invalid.

On the other hand, a symmetry breaking phase transition in the early
universe copiously produces solitonic monopoles 
with an abundance that would overdominate the present universe,
unless the symmetry breaking scale is very low.
This was one of the motivations for inflationary cosmology, which
dilutes away the monopoles by a period of a rapid cosmological
expansion~\cite{Guth:1980zm,Einhorn:1980ik,Linde:1981mu,Albrecht:1982wi}.
By solving the monopole problem, however, cosmic inflation also
prohibits relic monopoles from a phase transition to be observed.

Another possible venue for monopole production is in strong magnetic
fields, from which monopole-antimonopole pairs are non-perturbatively
produced.
This is the magnetic dual of the Schwinger
process~\cite{Sauter:1931zz,Heisenberg:1935qt,Schwinger:1951nm},
and the rate of monopole pair production with arbitrary coupling was
computed using an instanton method
in~\cite{Affleck:1981ag,Affleck:1981bma}. 
Monopole production from magnetic fields in 
magnetars (highly magnetized neutron stars with fields up to $B \sim
10^{15}\, \ro{G} \sim 10^{-5}\, \ro{GeV}^2$) and
heavy-ion collisions ($B \sim 10^{18}\, \ro{G} \sim 10^{-2}\, \ro{GeV}^2$
at CERN Super Proton Synchrotron) 
have been studied,
by further taking into account finite-temperature effects into the
calculations~\cite{Gould:2017fve,Gould:2017zwi,Gould:2018ovk,Gould:2019myj}
(see~\cite{Rajantie:2019oyc} for a review).
However even in such environments the magnetic fields are not strong
enough to produce monopoles with masses much larger
than a~$\ro{GeV}$. 

Magnetic fields also exist in the cosmic space on various
length scales, with their origin still remaining a mystery.
Spiral galaxies are known to host magnetic fields of
$ B \sim 10^{-5} \, \ro{G}$~\cite{Widrow:2002ud}.
Recent gamma ray observations suggest the presence of magnetic
fields even in intergalactic voids with strength
$B \gtrsim 10^{-15}\, \ro{G}$ coherent on Mpc scales or
larger~\cite{Tavecchio:2010mk,Neronov:1900zz,Dermer:2010mm}, 
giving strong indication that they are remnants of primordial magnetic
fields produced in the early universe.
Importantly, if (some parts of) the cosmological magnetic fields are
actually of primordial origin, then even if their present field strengths
are weak, they could have been extremely strong in the early universe.

In this paper we show that (even superheavy) monopole-antimonopole pairs
can be produced by primordial magnetic fields, and explore their
cosmological implications.
A strong enough primordial magnetic field dissipates energy
by the monopole pair production, and also by accelerating the monopoles.
By evaluating these effects,
we obtain consistency conditions for primordial magnetic fields to
survive until today to make up the observed magnetic fields, 
within physical theories that contain either elementary or solitonic
monopoles. 
We also discuss the possibility of primordial magnetic fields producing
an observable abundance of monopoles in the universe,
or even giving rise to a new kind of monopole problem.
Based on these discussions, we derive lower bounds on the monopole
mass, under the assumption that the observed cosmological magnetic
fields have a primordial origin.

Our discussion regarding the dissipation of primordial magnetic fields 
is quite distinct from those of the so-called ``Parker limit'' on the 
monopole flux, obtained by requiring the survival of galactic magnetic
fields~\cite{Parker:1970xv,Turner:1982ag}. 
(See also \cite{Long:2015cza} which studied the Parker limit for primordial
magnetic fields.)
While these works assume a hypothetical abundance of pre-existing
monopoles, 
here the monopoles are produced by the primordial magnetic field
itself and thus the monopole abundance is uniquely
determined.\footnote{If there are additional monopole producing
processes such as a thermal production~\cite{Turner:1982kh}, our bound
becomes tighter.} 
This enables us to obtain a direct bound on the monopole mass, as a
function of the primordial magnetic field strength.

This paper is organized as follows:
In Section~\ref{sec:production} we compute the number of monopoles
produced in primordial magnetic fields.
In Section~\ref{sec:effects} we evaluate the magnetic field dissipation
by the monopoles, as well as the monopole relic abundance, and derive
limits on the primordial magnetic field strength.
In Section~\ref{sec:limit},
the magnetic field limit is translated into limits on the monopole mass,
and the energy scale of magnetic field generation.
We then conclude in Section~\ref{sec:conc}. 
In Appendix~\ref{app:E} we present a general formalism for analyzing the
magnetic field dissipation by monopoles, and also analyze
effects that are not discussed in the main text.
In Appendix~\ref{app:reheating} we give the relations between the
Hubble rate, cosmic temperature, and redshift during the
reheating and radiation-dominated epochs.
In Appendix~\ref{app:PT} we give a lower limit on the relic abundance of
solitonic monopoles produced at a symmetry breaking phase transition. 

Throughout this paper we use Heaviside-Lorentz units,
with $c = \hbar = k_{\mathrm{B}} = 1$.
$M_{\mathrm{Pl}} $ refers to the reduced Planck mass~$ (8 \pi G)^{-1/2}$.
Unless explicitly noted, our discussions cover both
elementary and solitonic monopoles. 
The magnetic charge of the monopole is typically large 
(e.g. $g \approx 21 n$ for $e \approx 0.30$),
however most of the analyses apply even if $g$ is tiny.

\section{Monopole Pair Production in Magnetic Fields}
\label{sec:production}

\subsection{Vacuum Decay Rate}

Analyses of pair production in an external field
often invokes a weak coupling, 
as was also assumed by Schwinger~\cite{Schwinger:1951nm}, 
however this does not necessarily apply to monopoles due to the Dirac
quantization condition. 
Using an instanton method, 
the authors of~\cite{Affleck:1981ag,Affleck:1981bma}
derived an expression for the vacuum decay rate due to
monopole-antimonopole pair production in a
static magnetic field as
\begin{equation}
 \Gamma = \frac{(g B)^2}{(2 \pi)^3}
\exp \left[ -\frac{\pi m^2}{g B} + \frac{g^2}{4} \right],
\label{eq:Gamma}
\end{equation}
where $B$ is the magnetic field strength,
$m$ is the monopole mass, and $g$ is the amplitude of the magnetic
coupling (thus $g$ is non-negative hereafter).
This result is valid for an arbitrary~$g$, as long as the field
is sufficiently weak such that
\begin{equation}
 \frac{g B}{m^2} \lesssim 1,
\label{eq:0a}
\end{equation}
\begin{equation}
 \frac{g^3 B}{m^2} \lesssim 4 \pi .
\label{eq:0bp}
\end{equation}
The second condition suggests that the expression~(\ref{eq:Gamma}) is
valid while its exponent is negative, and this is stricter than the
first condition if $ g \gg 1$.
It can be understood as the requirement that, in order for the
semi-classical instanton techniques used to obtain (\ref{eq:Gamma}) to
be valid, the loop radius of the classical instanton
solution~\cite{Affleck:1981ag,Affleck:1981bma},
\begin{equation}
 R = \frac{m}{g B},
\label{eq:inst_rad}
\end{equation}
should be larger than the size of a monopole,\footnote{The classical
radius of a vanilla 't Hooft--Polyakov monopole is
of~(\ref{eq:mono-size})~\cite{Kirkman:1981ck}. It has been claimed that
elementary monopoles should also have a similar spatial
extension~\cite{Schiff:1966nps,Goebel:1970,Goldhaber:1983};
one simple argument is that the classical point-particle picture should
break down at distances shorter than (\ref{eq:mono-size}), since
otherwise the sum of the rest energy and potential energy of a
monopole-antimonopole pair can become negative and render the vacuum
unstable.}
\begin{equation}
 r \sim \frac{g^2}{4 \pi m}.
\label{eq:mono-size}
\end{equation}

The expression~(\ref{eq:Gamma}) can receive corrections also from
finite-temperature effects,
when the inverse of the temperature of the thermal bath is smaller than
the instanton radius, $1/T < R$; 
such thermal corrections to the monopole production rate have been
computed in~\cite{Gould:2017fve,Gould:2018ovk}. 
Gravitational effects on the monopole production are less studied, but
one naively expects that the rate receives corrections when 
the curvature radius of the spacetime is smaller than~$R$;
in a Friedmann--Robertson--Walker (FRW) universe, this condition is
written as $1/H < R$ where $H$ is the Hubble expansion
rate.\footnote{Pair production of charged scalar particles by
electric fields in de Sitter space was analyzed
in~\cite{Kobayashi:2014zza}. It was found that the flat-space
result is modified at $H^2 \gtrsim e E $, which is different from the
above naive guess of $1/H < R$ (with the replacement 
$g,B \rightarrow e,E$). 
It would be interesting to explicitly compute the monopole production
rate in a curved spacetime and check when gravitational effects become
important.}
Moreover, primordial magnetic fields redshift with the expansion of
the universe on a time scale of order the Hubble time;
this time dependence can also modify the rate if $1/H < R$
(see e.g. \cite{Gould:2019myj} for discussions on pair production in
spacetime-dependent fields).
These corrections would enhance the pair production rate,
even enabling the pair production to proceed via sphalerons, 
if the temperature and/or the Hubble scale are sufficiently high.

We should also note that in (\ref{eq:Gamma}), $m$ and $g$ should be
taken to be the renormalized quantities~\cite{Affleck:1981bma}. However
we ignore their runnings, which should be good enough for the
approximate calculations in this paper.

\subsection{Number Density}

In order to evaluate the number of monopoles produced from primordial
magnetic fields, we identify the decay rate~(\ref{eq:Gamma})
with the rate of pair production per unit volume per unit
time.\footnote{The two rates are not necessarily the
same~\cite{Cohen:2008wz}. The pair production rate in an 
electric field is computed in~\cite{Nikishov:1970br},
however it is also found that this matches with the vacuum decay rate in
the weak field limit.
Hence we suppose that they also match for monopoles in weak fields.}
Then the number density~$n$ of monopole-antimonopole pairs follows
\begin{equation}
 \dot{n} = - 3 H n + \Gamma ,
\end{equation}
where an overdot denotes a derivative with respect to
physical time~$t$,
and the Hubble rate is 
$H = \dot{a} / a$ in terms of the scale factor~$a$. 
Considering the magnetic field to be effectively homogeneous, 
this equation is integrated to yield
\begin{equation}
 n(t) = \frac{1}{a(t)^3}
\int^t_{t_i} dt' \, a(t')^3 \Gamma (t'),
\label{eq:n-int}
\end{equation}
where $t_i$ denotes the time when the magnetic field is switched on. 
Here, $\Gamma$ depends on time through its dependence on the 
magnetic field which redshifts with the expansion of the universe. 
We parameterize the redshifting of the magnetic field strength as
\begin{equation}
 B \propto a^{-p},
\label{eq:B-scaling}
\end{equation}
with $p$ being a positive constant of order unity. 
Without any source, primordial magnetic fields redshift with
$p = 2$, however different values of~$p$ can also be realized in the
presence of matter or with stronger electric
fields~\cite{Kobayashi:2019uqs}. 
Here, let us also introduce a dimensionless quantity
\begin{equation}
 \epsilon \equiv \frac{g B}{\pi m^2},
\label{eq:epsilon}
\end{equation}
which obeys $\epsilon \ll 1$ when the first weak field
condition~(\ref{eq:0a}) is well satisfied.
Since the production rate~$\Gamma$ depends exponentially on~$B$, 
it decays very quickly under weak fields on a time scale of
$\Delta t_\Gamma = \abs{\Gamma / \dot{\Gamma}} \simeq \epsilon / (p H) \ll
1 / H$. 
Hence the integral in (\ref{eq:n-int}) is dominated by the lower limit,
and we obtain an approximate expression for the pair number density
valid for $ t \gtrsim t_i + \Delta t_{\Gamma i}$ as\footnote{The integral
in (\ref{eq:n-int}) can be directly 
performed when the background universe has a constant equation of
state~$w$ such that $H \propto a^{-3 (1+w)/2}$, as
\begin{equation}
 n = e^\frac{g^2}{4}
\frac{1}{ 8 \pi p} \frac{m^4}{H_i} 
\left( \frac{a_i}{a} \right)^3 
\epsilon_i^{b + 2} 
\left[
\mathcal{G} \left( b ,  \frac{1}{\epsilon_i} \right) -
\mathcal{G} \left( b ,  \frac{1}{\epsilon} \right)
\right],
\quad \mathrm{where} \quad
b = \frac{3 (3+w)}{2p} - 2.
\end{equation}
Here we assumed $b > 0$, and 
$\mathcal{G}(b,z) = \int_z^\infty x^{b-1} e^{-x} dx $
is the incomplete gamma function.
By using the asymptotic form 
$\mathcal{G} (b, 1/\epsilon)  \sim \epsilon^{-b+1}
e^{-1/\epsilon } $ in the weak field limit $\epsilon \to 0$,
one obtains 
\begin{equation}
 n \sim 
\frac{\epsilon_i \Gamma_i}{p H_i} 
\left( \frac{a_i}{a} \right)^3 -
\frac{\epsilon \Gamma}{p H}.
\end{equation}
The first term dominates at $ t \gtrsim t_i + \Delta t_{\Gamma i}$,
and then the expression reduces to (\ref{eq:n-approx}).}
\begin{equation}
n \sim 
\frac{(\Delta t_{\Gamma} \, a^3 \Gamma )_i }{a^3}
= 
\frac{\epsilon_i \Gamma_i }{p H_i}
\left( \frac{a_i}{a} \right)^3,
\label{eq:n-approx}
\end{equation}
where quantities measured at the initial time~$t_i$
is denoted by the subscript~$i$.

The ``initial time''~$t_i$ which we defined
as the moment when the magnetic field switches on and begins to redshift
as~(\ref{eq:B-scaling}), can be understood as 
the time when the magnetic field generation has concluded.
To keep our discussion general we do not specify the concrete mechanism
of primordial magnetic field generation, but the time when the
generation process completes could for instance be at the end of
inflation~\cite{Turner:1987bw,Ratra:1991bn}, after inflation when the
universe is dominated by an oscillating
inflaton~\cite{Kobayashi:2014sga}, or at cosmological phase
transitions~\cite{Vachaspati:1991nm,Cornwall:1997ms}.

\section{Effects of Produced Monopoles}
\label{sec:effects}

As one can read off from the expression~(\ref{eq:Gamma}) for $\Gamma$, 
the exponential suppression factor disappears and 
monopole production becomes significant as
the magnetic field strength approaches the value,
\begin{equation}
 B_\star  = 4 \pi \frac{m^2}{g^3}.
\label{eq:B_up}
\end{equation}
This is also the field strength which saturates the second weak field
condition~(\ref{eq:0bp}). In this section we show that primordial
magnetic fields could not have been stronger than $B_{\star}$, by
evaluating the backreaction of the monopoles on the magnetic field, 
and also the monopole relic abundance.

We will mostly consider times long before the electroweak phase
transition, therefore the primordial magnetic field and the monopoles
are actually those of the hypercharge U(1) gauge field.
When these are converted into the magnetic fields and monopoles of the
electromagnetic U(1) at the electroweak phase transition, 
quantities such as the magnetic field strength and 
magnetic charge will change by 
a number of order unity that depends on the Weinberg angle,
however this will not be important for our discussions.

\subsection{Magnetic Field Dissipation by Monopole Production}

In terms of the energy density of the magnetic field,
\begin{equation}
 \rho_B = \frac{B^2}{2},
\end{equation}
the magnetic field scaling~(\ref{eq:B-scaling}) is rewritten as
\begin{equation}
 (\dot{\rho}_B)_{\mathrm{red}} = - 2 p H \rho_B,
\label{eq:Gamma_B}
\end{equation}
where we have added a subscript ``red'' to specify that this
contribution to $\dot{\rho}_B$ represents the redshifting of the
magnetic field due to the cosmic expansion.
The time scale of redshifting,
$\Delta t_{\ro{red}} = \abs{\rho_B / (\dot{\rho}_B)_{\ro{red}}} = 1/2 p H$,
is of order the Hubble time.

Additionally, each time the field produces a monopole-antimonopole pair
it looses energy corresponding to the rest energy of the
pair, $ \Delta \mathcal{E}_B = - 2 m $.\footnote{By equating the pair's
rest energy with its potential 
energy due to the background magnetic field, one obtains the critical
separation between the pair upon creation as $ r_c = 2 m / g B$, which
is of the same order as the instanton radius~(\ref{eq:inst_rad}). 
The depletion of the field energy $ \Delta \mathcal{E}_B = - 2 m $
corresponds to the decrease in the net magnetic field
strength due to a monopole-antimonopole pair separated by~$r_c$.
Under the weak field condition~(\ref{eq:0bp}),
the attractive force between a pair separated by~$r_c$
is weaker than the repulsive force imposed by the background field.}
Thus the energy dissipation due to pair production per unit time and
volume is 
\begin{equation}
 (\dot{\rho}_B)_{\mathrm{prod}} = - 2 m \Gamma.
\label{eq:Gamma_m}
\end{equation}
This dissipation rate is smaller than the rate of
redshifting, i.e. $2 m \Gamma < 2 p H \rho_B$, if
\begin{equation}
 B < B_{\mathrm{prod}} = 
\frac{4 \pi m^2}{g^3}
\left[ 1 + 
\frac{4}{g^2} \ln \left( \frac{g^2}{4 \pi^3 p} \frac{m}{H} \right)
\right]^{-1} .
\label{eq:B_m}
\end{equation}
A magnetic field stronger than the right hand side
would quickly decay through monopole production on a time scale
shorter than a Hubble time, until the field falls
below~$B_{\mathrm{prod}}$.  
Hence (\ref{eq:B_m}) gives an upper bound on the primordial magnetic field
strength.

The expression for~$B_{\mathrm{prod}}$ becomes negative if 
$(4/g^2) \ln [(g^2 / 4 \pi^3 p) (m/H)] < -1$.
This implies that in such a case the dissipation becomes
significant only at very strong fields where the
expression~(\ref{eq:Gamma}) for~$\Gamma$ breaks down.
However, the logarithmic term is generically larger than~$-1$ 
if $ g \gg 1$:
Even in the extreme case where the monopole mass saturates its lower
bound $m \gtrsim 1\, \ro{GeV}$ derived from 
heavy-ion collisions~\cite{Gould:2017zwi},\footnote{A similar 
bound on~$m$ can also be derived by combining discussions on the thermal
production of monopoles~\cite{Turner:1982kh} with the lower limit on the
reheating temperature from Big Bang Nucleosynthesis (BBN).}
and the Hubble scale saturates the upper bound 
$H \lesssim 10^{14}\, \ro{GeV}$ on the inflation
scale~\cite{Aghanim:2018eyx},
we get $(4/g^2) \ln (m/H) \gtrsim -0.5$ as long as $g \gtrsim 16$.
Hence by supposing $ g \gg 1$ and considering the logarithmic term to be
either negligible or positive, 
we get $B_{\mathrm{prod}} \lesssim B_{\star}$,
which guarantees that the weak field conditions
(\ref{eq:0a}) and (\ref{eq:0bp}) are satisfied at 
$B = B_{\mathrm{prod}}$.

One can also check that, under the weak field conditions,
the ratio $B/B_{\ro{prod}}$ monotonically decreases in time,
which indicates that the energy dissipation by the monopoles 
is more important at earlier times.
Hence (\ref{eq:B_m}) also sets a lower limit on the time
of magnetic field generation, as we will see explicitly in 
Section~\ref{sec:limit}.

\subsection{Magnetic Field Dissipation by Monopole Acceleration}

After the monopoles are produced, they are accelerated by the magnetic
fields and thus further deplete the magnetic field energy.\footnote{This
discussion of monopole acceleration follows that
of~\cite{Parker:1970xv,Turner:1982ag}. There is, however, a key
difference that here the monopole abundance is produced by the
magnetic field itself and thus is uniquely determined.}
We first evaluate this effect by assuming 
that the (anti)monopoles move with
relativistic velocities $v \simeq 1$ in the (reverse) direction of the
magnetic field.
Then each (anti)monopole gains kinetic energy 
of $\Delta \mathcal{E}_M = g B \Delta t$,
and in turn the magnetic field loses energy per unit time and volume as
\begin{equation}
 (\dot{\rho}_B)_{\mathrm{R}} = - 2 n g B.
\label{eq:Gamma_dis}
\end{equation}
Let us for the moment only consider pairs that
are produced during an interval~$\Delta t_\Gamma$ 
around the time of consideration,
and substitute for the pair density
(see discussions around (\ref{eq:n-approx})),
\begin{equation}
n \to  \frac{\epsilon \Gamma }{p H}.
\label{eq:n-at-the-time}
\end{equation}
This amounts to ignoring energy dissipation by the accumulated abundance
of monopoles produced in the past, and thus we will obtain a
conservative bound on the field strength.
Then one finds that the dissipation rate~(\ref{eq:Gamma_dis}) due to
accelerating relativistic monopoles, is smaller than the rate of
redshifting~(\ref{eq:Gamma_B}),
i.e. $2 n g B < 2 p H \rho_B$, when 
\begin{equation}
 B < B_{\mathrm{R}} = 
\frac{\pi m^2}{2 g}
\frac{1}{W(x_{\ro{R}})}
\quad \ro{with} \quad
x_{\ro{R}} = e^{\frac{g^2}{8}} \frac{g }{4 \pi p}  \frac{m}{H} .
\label{eq:B_dis}
\end{equation}
Here, $W(x)$ is the Lambert $W$-function which is a solution of $W
e^W=x$; it is non-negative and increasing for $x \geq 0$. 
$B_{\mathrm{R}}$ also serves as an upper limit on the magnetic field
strength, beyond which the field quickly decays by accelerating
the monopoles the field itself has produced.

One can obtain an approximate expression for $B_{\mathrm{R}}$
if the first weak field condition~(\ref{eq:0a}) is well satisfied, 
i.e. $2 g B_{\mathrm{R}} / \pi m^2 = 1/ W (x_{\ro{R}}) \ll 1$.
This is equivalent to $x_{\ro{R}} \gg 1$,
for which we can use the rough approximation
$W(x_{\ro{R}}) \sim \ln x_{\ro{R}}$~\cite{NIST} to obtain
\begin{equation}
B_{\mathrm{R}} \sim 
 \frac{4 \pi m^2}{g^3}
\left[ 1 + 
\frac{8}{g^2} \ln \left( \frac{g }{4 \pi p} \frac{m}{H} \right)
\right]^{-1}.
\label{eq:B_dis-approx}
\end{equation}
This now takes a form similar to the upper bound~(\ref{eq:B_m}) from
pair production, except for the logarithmic factor.

In the above discussions we assumed the monopoles to be relativistic,
however one obtains similar results also for non-relativistic monopoles.
From the equation of motion of a non-relativistic monopole/antimonopole,
$m \ddot{z} = \pm g B$ where $z$ is the direction of the magnetic field,
the distance the monopole/antimonopole travels is 
$\Delta z = \pm (g B / 2 m) (\Delta t)^2$,
supposing they are initially at rest.
Here we ignored the time dependence of~$B$ as well as the effect of the
cosmological expansion on the monopole dynamics;
this is because 
we are interested in cases where the magnetic field is
dissipated on time scales comparable to or shorter than a Hubble time,
and also because we use this computation only while the magnetic field
strength changes by an order-unity factor.\footnote{When taking into
account the backreaction of the monopoles, one obtains an oscillatory
solution for the magnetic field; this magnetic field oscillation is
expected to decay by a Landau damping~\cite{Turner:1982ag}.}
Thus by accelerating $n$~pairs from zero initial velocity,
the magnetic field looses its energy density during $\Delta t$ as
\begin{equation}
 - \Delta \rho_B = 
2 n g B \abs{\Delta z} 
= \frac{n g^2 B^2 }{m} (\Delta t)^2.
\end{equation}
Equating this with $\rho_B$, one obtains the characteristic time scale
of magnetic energy dissipation,
\begin{equation}
 \Delta t_{\mathrm{NR}} = \frac{1}{g} 
\sqrt{ \frac{m}{2 n} }.
\label{eq:t_dis}
\end{equation}
The condition for this to be longer than the time scale of redshifting
$\Delta t_{\ro{red}} = 1 / 2 p H$ (cf. (\ref{eq:Gamma_B}))
is obtained by substituting (\ref{eq:n-at-the-time}) for~$n$ as
\begin{equation}
 B < B_{\ro{NR}} = 
\frac{\pi m^2}{3 g} \frac{1}{W(x_{\ro{NR}} )}
\quad \ro{with} \quad
x_{\ro{NR}} = e^{\frac{g^2}{12}} 
\frac{g^{2/3} }{ 6 (2 \pi)^{1/3} p}  \frac{m}{H} .
\label{eq:B_dis-NR}
\end{equation}
Under the first weak field condition~(\ref{eq:0a}), this upper limit is
approximated by 
\begin{equation}
B_{\ro{NR}}
\sim 
 \frac{4 \pi m^2}{g^3}
\left[ 1 + 
\frac{12}{g^2} \ln 
\left( \frac{g^{2/3} }{ 6 (2 \pi)^{1/3} p}  \frac{m}{H}  \right)
\right]^{-1}.
\label{eq:B_dis-NR-approx}
\end{equation}
Since (\ref{eq:B_dis-approx}) and (\ref{eq:B_dis-NR-approx}) differ only
by the logarithmic factor, we conclude that the order of magnitude of
the  magnetic field limit does not depend on whether the monopoles are
relativistic or not. 

We have treated the primordial magnetic field as
effectively homogeneous, by supposing the coherence length of the
field to be larger than the Hubble radius at the
time of consideration. 
In fact, the observationally hinted intergalactic magnetic field
typically has a coherence length of Mpc scale or
larger~\cite{Tavecchio:2010mk,Neronov:1900zz,Dermer:2010mm}, which,
if it is of primordial origin, re-enters the horizon only at $a_0/a
\lesssim 10^6$. However if the primordial magnetic field had
inhomogeneous components with sub-horizon coherence lengths 
in the early universe, then the monopoles would not always
travel in the direction of the magnetic field and thus the energy
dissipation via monopole acceleration would be less effective
(see~\cite{Turner:1982ag} for similar discussions for monopoles in
galactic magnetic fields).

We also remark that, in the above
analyses we only included monopoles produced ``on the spot,'' and
ignored monopoles that have already been produced in the past. 
Depending on the time evolution of the monopole velocity, the
population of monopoles from the past may more effectively deplete the
magnetic field energy, in which case the bound on the magnetic
field strength becomes even tighter. 
Discussions on this point, as well as a general formalism for analyzing
the magnetic field dissipation by both the production and acceleration
of all existing monopoles, are presented in Appendix~\ref{app:E}.

\subsection{Monopole Relic Abundance and Flux}
\label{sec:abundance}

Constraints on the relic density of the produced monopoles yield 
further limits on primordial magnetic fields.
Supposing the monopoles today to be non-relativistic, 
their relic density is obtained as
\begin{equation}
 \rho_{M 0} = 2 m n_0 \sim 
\frac{2 m \epsilon_i \Gamma_i }{p H_i}
\left( \frac{a_i}{a_0} \right)^3,
\label{eq:rho_M0}
\end{equation}
where we used (\ref{eq:n-approx}),
and the subscript ``$0$'' denotes values in the present universe.
Requiring the density of monopoles not to exceed that of dark matter, 
i.e. $ \rho_{M0}  < \rho_{\ro{dm} 0} \approx 0.3 \rho_{\ro{crit}
0}$~\cite{Aghanim:2018eyx}, 
we obtain an upper limit on the initial magnetic field strength 
(the value when magnetic field generation concludes) as
\begin{equation}
 B_i < B_{\ro{dm}} = 
\frac{\pi m^2}{3 g}
\frac{1}{W(x_{\ro{dm}})}
\quad \ro{with} \quad
x_{\ro{dm}} = e^{\frac{g^2}{12}} 
\left( \frac{1}{108 \pi p} \frac{m^5}{H_i \rho_{\ro{dm} 0}} \right)^{1/3} 
\frac{a_i}{a_0} . 
\label{eq:B_xi}
\end{equation}

In order to evaluate $a_i / a_0$,
let us suppose that the generation of the primordial magnetic field 
concludes at the end of inflation or later, but before
matter-radiation equality. 
We further assume a post-inflation history starting with 
an epoch dominated by an oscillating inflaton,
which eventually decays away and initiates the
radiation-dominated epoch.
We use the subscript ``end'' to denote quantities at the end of
inflation, ``dom'' at the time when radiation domination takes over,
and ``eq'' at matter-radiation equality.

The Hubble scale during the post-inflation epochs as a function of the
scale factor is given in Appendix~\ref{app:reheating}.
Using (\ref{eq:coll-H}) and (\ref{eq:adom-Tdom}) to rewrite $a_i$ in
terms of $H_i$, one obtains 
\begin{equation}
 x_{\ro{dm}} \sim e^{\frac{g^2}{12}} \frac{1}{p^{1/3}}
\left( \frac{m}{10^{9}\, \ro{GeV}} \right)^{5/6}
\left( \frac{m}{H_i} \right)^{5/6}
\ro{min.}
\left\{
1, \left( \frac{H_{\ro{dom}}}{H_i} \right)^{1/6}
\right\},
\end{equation}
where the last factor depends on whether the magnetic field generation
completes during radiation domination ($H_i < H_{\ro{dom}}$),
or in an earlier epoch ($H_i > H_{\ro{dom}}$).
Further using the weak field condition, the magnetic field limit is thus
approximately written as
\begin{equation}
B_{\ro{dm}}
\sim 
 \frac{4 \pi m^2}{g^3}
\left[ 1 + 
\frac{12}{g^2} \ln 
\left( 
\frac{1}{p^{1/3}}
\left( \frac{m}{10^{9}\, \ro{GeV}} \right)^{5/6}
\left( \frac{m}{H_i} \right)^{5/6}
\ro{min.}
\left\{
1, \left( \frac{H_{\ro{dom}}}{H_i} \right)^{1/6}
\right\}
\right)
\right]^{-1},
\label{eq:B_xi-approx}
\end{equation}
which differs from the other bounds only by the logarithmic factor.

A few comments are in order. 
First, since the monopoles are continuously accelerated by the magnetic
fields, they may withstand the Hubble damping and be moving with
relativistic velocities in the current universe 
(cf. Appendix~\ref{app:velocity}).
In such a case, the relic density is larger than~(\ref{eq:rho_M0}),
and the relativistic monopoles serve as extra radiation 
which is constrained by CMB observations and BBN,
yielding a tighter bound on~$B_i$.
Secondly, we assumed that the annihilation of monopoles and
antimonopoles does not significantly reduce their abundance.
According to the analyses in~\cite{Zeldovich:1978wj,Preskill:1979zi},
the annihilation only becomes relevant if the monopole number density is
so large as to lead to an overabundance (unless the mass is very light).
The discussion may be modified in the presence of cosmological magnetic
fields, which pull the monopoles and antimonopoles apart. It will be
interesting to study annihilation effects in a magnetic field background. 

One can further compute the average flux of monopoles,
$F = 2 n_0 v_0 / 4 \pi$,
with $v_0$ being the monopole velocity,
and compare with various existing
bounds~\cite{Mavromatos:2020gwk,Zyla:2020zbs} including the  
Parker limit~\cite{Parker:1970xv,Turner:1982ag,Long:2015cza}.
Depending on the monopole mass, the flux bounds give stronger
constraints than $\rho_{M 0 } < \rho_{\ro{dm} 0}$.
However they do not drastically improve the 
limit on~$B_i$, which depends on the bound on the monopole
abundance only through the logarithmic factor.

\subsection{Remarks on Solitonic Monopoles}
\label{sec:remarks}

For monopoles that are topological solitons of spontaneously broken
gauge theories, the discussions above do not apply when the symmetry is
unbroken, as then the monopole solution does not exist. 
Thus strong magnetic fields can exist without producing monopoles
while the cosmic temperature and/or the Hubble rate is larger than the
symmetry breaking scale, $\sigma < \ro{max.}\{ T, H \}$.
Even with a low temperature and Hubble rate, 
the magnetic field itself may restore the symmetry if it is stronger
than~$B_{\star}$~\cite{Salam:1974xe,Kirzhnits:1976ts,Shore:1981mj}.
(The explosive production of monopole-antimonopole pairs at 
$B \sim B_{\star}$ may be related to this symmetry restoration.)
However if the symmetry is unbroken at some time during the
post-inflation era, then later at the symmetry breaking phase transition,
monopoles are copiously produced and eventually overdominate the
universe, unless the symmetry breaking scale is very low.
It should also be noted that this monopole problem is particularly severe 
if the phase transition happens prior to radiation domination
(see Appendix~\ref{app:PT}).

Hence, although the magnetic field limits can in principle be evaded by
keeping the symmetry unbroken, one has to pay the price of endangering
the universe with the monopole problem. 
Therefore it is safe to assume also for solitonic monopoles that
primordial magnetic fields beyond the aforementioned upper limits
could not have existed in the post-inflation universe.

We also remark that in a phase of broken symmetry, the magnetic field
strength in a radiation-dominated universe is bounded from above by the
symmetry breaking scale, thus is typically well below~$B_{\star}$. 
On the other hand, stronger fields can exist in the
pre-radiation-dominated era, during which 
solitonic monopoles can be abundantly produced.
We will see this in detail in Section~\ref{sec:f_limit}.

\subsection{Summary: A Conservative Bound}
\label{sec:summary}

In the previous subsections we derived upper limits on the 
primordial magnetic field amplitude, 
beyond which the magnetic field self-screens
by producing monopole-antimonopole pairs~(\ref{eq:B_m}),
or by accelerating the produced monopoles~(\ref{eq:B_dis-approx}),
(\ref{eq:B_dis-NR-approx}). 
A cosmological upper limit~(\ref{eq:B_xi-approx}) was also derived by
the requirement 
that the magnetic field does not overproduce monopoles in the universe. 
These limits from self-screening and overproduction are
all comparable to or smaller than~$B_{\star}$, as one can check by
following a discussion similar to that below~(\ref{eq:B_m}).

Potential loopholes to the individual limits were discussed in each
subsection, but let us give a few more remarks:
\begin{itemize}
 \item {\it Corrections to pair production rate.} 
The expression~(\ref{eq:Gamma}) for~$\Gamma$ used in our analyses can
       break down at the values of~$B$ where self-screening or monopole
       overproduction takes place, if (a) the weak field conditions are
       violated, or (b) the cosmic temperature/Hubble scale are
       sufficiently high to induce finite-temperature/gravitational
       corrections. We discussed below (\ref{eq:B_m}) that (a) is
       unlikely for $g \gg 1$, however there can still be some corrections
       to~$\Gamma$ since $B_{\star}$ only marginally satisfies the
       second weak field condition~(\ref{eq:0bp}). Regarding (b), note
       that the radius~(\ref{eq:inst_rad}) of the classical instanton
       solution at 
       $B_{\star}$ is $R_{\star} = g^2 / (4 \pi m)$. For solitonic
       monopoles whose masses are related to the symmetry breaking scale
       typically via $m \sim g \sigma$, one finds during the symmetry
       broken phase, i.e. $\sigma >  T, H $, that $(4 \pi / g) R_{\star}
       \lesssim  1/T, 1/H $. If $ g \sim 10$ we get $R_{\star} \lesssim
       1/T, 1/H $, and hence we can safely use the zero-temperature and
       flat-space expression~(\ref{eq:Gamma}). 
       The corrections, however, may become important if the actual
       limits such as~$B_{\ro{prod}}$ are much smaller than~$B_{\star}$,
       or in the late universe when $B$ is small.
       We also note that this discussion based on the
       symmetry breaking scale does not directly apply to elementary
       monopoles. When (\ref{eq:Gamma}) breaks down, pair
       production tends to take place at a faster rate. Therefore we
       expect that corrections to~$\Gamma$, if any, can only make the upper
       limits on the primordial magnetic field more stringent.

 \item {\it Interaction with thermal plasma.} 
We have ignored the interaction of monopoles with the thermal plasma,
which can affect our discussions in the following ways: (a) The friction
from the plasma may slow down the monopole (see
e.g.~\cite{Meyer-Vernet:1985yyn}) and render the magnetic field
dissipation via monopole acceleration less efficient.
(b) Strong magnetic fields in the pre-radiation-dominated era 
(such as those generated in the magnetogenesis scenarios
of~\cite{Turner:1987bw,Ratra:1991bn,Kobayashi:2014sga}) 
can give away large energy to the monopoles, which in turn may raise the
temperature of the plasma. If such a monopole-mediated reheating were to
happen, it would modify the perturbative reheating history we assumed
for evaluating the monopole relic density.

 \item {\it Thermal production.} 
Monopoles can also be thermally produced in the early universe. In
       particular for solitonic monopoles, according to the analysis
       in~\cite{Turner:1982kh}, there can be a temperature window below
       the symmetry breaking scale where an observable monopole
       abundance is thermally produced.
       However this analysis assumes entropy conservation after
       the monopole production, and thus is modified in the
       pre-radiation-dominated epoch. In any case, the existence of such
       an additional monopole population would further tighten the
       magnetic field limits we discussed. 

 \item {\it Effects on magnetic field generation.} 
Our field limits should be applied to primordial magnetic fields after
their generation process has completed. 
This is because during the magnetic field generation,
the field can grow faster than it is dissipated by the 
monopoles, and/or the U(1) gauge theory itself is modified such that the
magnetic energy density does not take the form $\rho_B = B^2 / 2$
(as is typically the case for Weyl symmetry-breaking magnetogenesis
       scenarios).  
If the monopoles are solitonic, modifications of the gauge theory
can further affect the monopole solution itself.
We also note that the calculation of the monopole relic density is
       modified if the magnetic field generation completes before the
       end of inflation. 
It would be interesting to study how monopoles affect various
magnetic field generating mechanisms.
\end{itemize}

While most of the effects discussed here and in each subsection 
further tighten our magnetic field limits,
some of them may weaken the limits.
However we also note that, none of the effects 
seem capable of evading all limits in one go.
Thus we conclude that if either elementary or solitonic monopoles are
contained in the physical theory, 
then the amplitude of primordial magnetic fields in the
post-inflation universe is always bounded from above as
\begin{equation}
 B \lesssim B_{\star} = 4 \pi \frac{m^2}{g^3}. 
\label{eq:smaller-than-B_up}
\end{equation}
We stress that this is a conservative upper bound, 
and the magnetic self-screening and/or monopole overproduction can
happen with weaker fields.

\section{Limits on Monopole Mass and Primordial Magnetic Field}
\label{sec:limit}

\subsection{General Limits}

We now discuss the implications of the
bound~(\ref{eq:smaller-than-B_up}) for primordial magnetic
field generation and monopoles.
Below we suppose the magnetic field to 
redshift consistently as $B \propto a^{-2}$,
i.e. (\ref{eq:B-scaling}) with $p = 2$, after being generated. 
Then $B_i < B_{\star}$ imposes a lower bound on the scale
factor (or equivalently an upper bound on the redshift) when the
magnetic field generation completes,
\begin{equation}
 a_i > a_{\star} = 
a_0 \left( \frac{B_0}{B_{\star}} \right)^{1/2},
\label{eq:a_up}
\end{equation}
with the right hand side expressed in terms of the present-day
magnetic field strength~$B_0$. 
This in turn sets an upper bound on the Hubble scale as $H_i < H(a_\star)$.
(We remind the reader that $H_i$ is the Hubble scale at the completion
of the magnetic field generation; 
see discussions below (\ref{eq:n-approx}). 
For instance in inflationary magnetogenesis scenarios where the magnetic
fields are excited during the inflation epoch, $H_i$ is equal to the
Hubble scale at the end of inflation, $H_{\rm{end}}$.)

We assume the magnetic field generation was completed either at the end
of inflation, or during the subsequent reheating or radiation-dominated
epochs (i.e. $t_{\ro{end}} \leq t_i < t_{\ro{eq}}$),
and adopt the usual post-inflation history based on perturbative
reheating as described in Section~\ref{sec:abundance} or
Appendix~\ref{app:reheating}. 
Then calculating $H(a_\star)$ using (\ref{eq:coll-H}) and
(\ref{eq:adom-Tdom}), 
we obtain an upper limit on~$H_i$, whose form depends on whether
$a_\star$ is smaller or larger than the scale factor upon radiation
domination~$a_{\ro{dom}}$,\footnote{Since $a_{\star} / a_0 \sim 10^{-29}
(B_0 / 10^{-15}\, \ro{G})^{1/2} (m g^{-3/2} / 10^{11}\, \ro{GeV})^{-1}$,
we can safely assume that $a_{\star} \ll a_{\ro{eq}} \approx a_0 / 3000$. 
On the other hand, depending on the inflation scale, $a_\star$ can even
be smaller than $a_{\ro{end}}$. In such cases our $H(a_{\star})$, 
obtained using (\ref{eq:coll-H}) which is valid for $t_{\ro{end}}
< t < t_{\ro{eq}}$, gives a conservative upper limit on~$H_i$.}
\begin{equation}\label{eq:Hi-limit}
\begin{split}
 & \quad
H_i \lesssim H_{\ro{dom}}
\, \, \ro{min.}\left\{
\left( \frac{a_{\ro{dom}}}{a_{\star}} \right)^{2},
\left( \frac{a_{\ro{dom}}}{a_{\star}} \right)^{3/2}
\right\},
\\
\frac{a_{\ro{dom}}}{a_{\star}} & \sim
\left( \frac{H_{\ro{dom}}}{10^{14}\, \ro{GeV}} \right)^{-1/2}
\left(\frac{B_0}{10^{-15}\, \ro{G}} \right)^{-1/2}
\left(\frac{m g^{-3/2}}{10^{11}\, \ro{GeV}} \right).
\end{split}
\end{equation}
The combination~$m g^{-3/2}$ derives from (\ref{eq:smaller-than-B_up}),
which traces back to the ratio between the two terms in the exponent
of~$\Gamma$, cf.~(\ref{eq:Gamma}). 
This upper limit on~$H_i$ can also be written as a lower limit on the
monopole mass, 
\begin{equation}
\frac{m}{g^{3/2}} \gtrsim 10^{11}\, \ro{GeV}
\left( \frac{B_0}{10^{-15}\, \ro{G}} \right)^{1/2}
\left( \frac{H_i}{10^{14}\, \ro{GeV}} \right)^{1/2}
\ro{max.} \left\{
1, \left( \frac{H_i}{H_{\ro{dom}}} \right)^{1/6}
\right\}.
\label{eq:m-limit}
\end{equation}
Thus we have obtained a consistency bound on monopoles~($m$, $g$) and
primordial magnetic fields~($B_0$, $H_i$), 
for a given post-inflation history characterized by~$H_{\ro{dom}}$.

The temperature at the onset of radiation domination, $T_{\ro{dom}}$, 
which is often referred to as the reheat temperature, 
is related to $H_{\ro{dom}}$ via~(\ref{eq:adom-Tdom}). 
It is required to lie within the range
$10^{-3}\, \ro{GeV} \lesssim T_{\ro{dom}} \lesssim 10^{16}\, \ro{GeV} $,
where the lower bound comes from BBN
and the upper bound is from the observational limit 
on the energy scale of inflation~\cite{Aghanim:2018eyx}.
The reference value for the present-day magnetic field in the above
expressions is taken from the claimed lower limit 
$B_0 \gtrsim 10^{-15}\, \ro{G}$ on intergalactic magnetic fields
from gamma ray
observations~\cite{Tavecchio:2010mk,Neronov:1900zz,Dermer:2010mm}. 
(If the coherence length~$\lambda$ of the magnetic field is much
smaller than a Mpc, the lower limit improves as~$\lambda^{-1/2}$.) 
We also remark that a primordial magnetic field, if homogeneous, 
is bounded from above as $B_0 \lesssim 10^{-9}\, \ro{G}$ from CMB
anisotropies~\cite{Barrow:1997mj}, although it has also 
been claimed that this limit is relaxed in the presence of
free-streaming particles like neutrinos~\cite{Adamek:2011pr}.

\begin{figure}[t]
 \begin{center}
 \begin{center}
 \includegraphics[width=.46\linewidth]{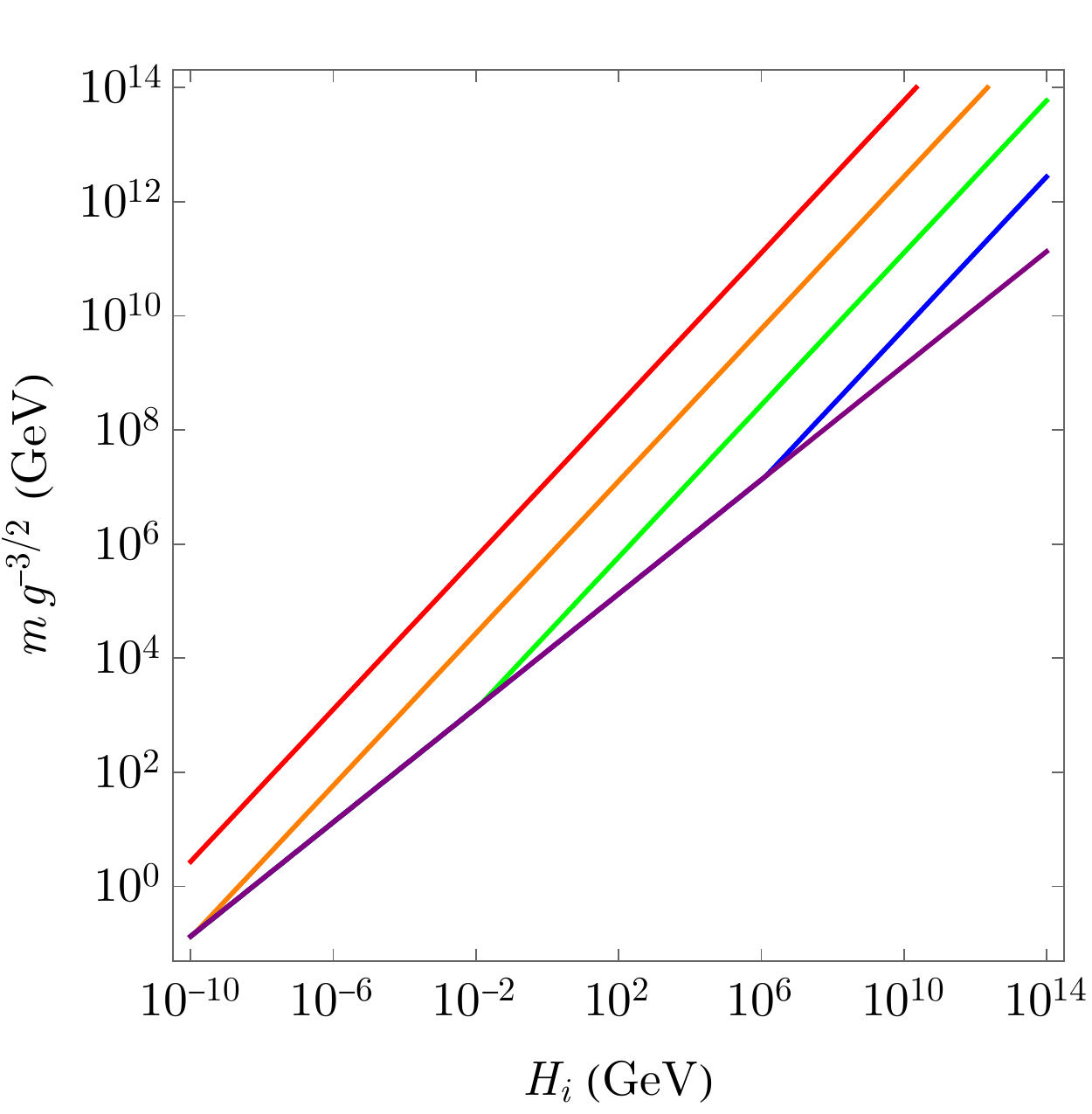}
 \end{center}
 \caption{\textbf{Lower limit} on $m g^{-3/2}$ 
 ($m$:  monopole mass, $g$: magnetic charge)
 as a function of the Hubble scale~$H_i$
 when primordial magnetic fields are initially generated.
 The present-day magnetic field strength is taken as
 $B_0 = 10^{-15}\, \ro{G}$. 
 The cosmic temperature when radiation domination takes over is varied as
 $T_{\ro{dom}} = 10^{16}\, \ro{GeV}$ (purple), $10^{12}\, \ro{GeV}$
 (blue), $10^{8}\, \ro{GeV}$ (green), $10^{4}\, \ro{GeV}$ (orange), and
 $1\, \ro{GeV}$ (red).}
 \label{fig:mg-Hi}
 \end{center}
\end{figure}

In Figure~\ref{fig:mg-Hi} we plot the lower limit (\ref{eq:m-limit}) on
$m g^{-3/2}$ as a function of~$H_i$
(or equivalently the upper limit~(\ref{eq:Hi-limit}) on~$H_i$ in terms
of~$m g^{-3/2}$).
Here, the present-day magnetic field strength is fixed to the minimum
value for intergalactic magnetic fields, 
$B_0 = 10^{-15}\, \ro{G}$,
and the limits are shown for 
$T_{\ro{dom}} = 10^{16}\, \ro{GeV}$ (purple), 
$10^{12}\, \ro{GeV}$ (blue),
$10^{8}\, \ro{GeV}$ (green),
$10^{4}\, \ro{GeV}$ (orange),
and $1\, \ro{GeV}$ (red).
The colored lines in the plot overlap at $H_i \leq H_{\ro{dom}}$,
where the limit becomes independent of $H_{\ro{dom}}$;
in other words, the bend in the line is at $H_i = H_{\ro{dom}}$. 
As one goes towards larger~$H_i$ (magnetic field generation at earlier
times), a stronger initial magnetic field is required to survive the
more substantial redshifting, and therefore the lower limit on 
$m g^{-3/2}$ becomes more stringent. 
The same is true for lower~$T_{\ro{dom}}$ when $H_i > H_{\ro{dom}}$,
which can be understood from the fact that the universe expands more
rapidly during an inflaton domination than radiation domination.

In the parameter regions slightly above the colored lines, 
an observable abundance of monopoles, but not so large as to
overdominate the universe, could be produced. One sees that even
monopoles of GUT scale mass
($m = 10^{16}\, \ro{GeV}$ with, say, $g = 10$ gives 
$m g^{-3/2} \sim 10^{14}\, \ro{GeV}$) 
are produced if magnetic field generation takes place at sufficiently
high energy scales.

\begin{figure}[t]
  \begin{center}
  \begin{center}
  \includegraphics[width=0.45\linewidth]{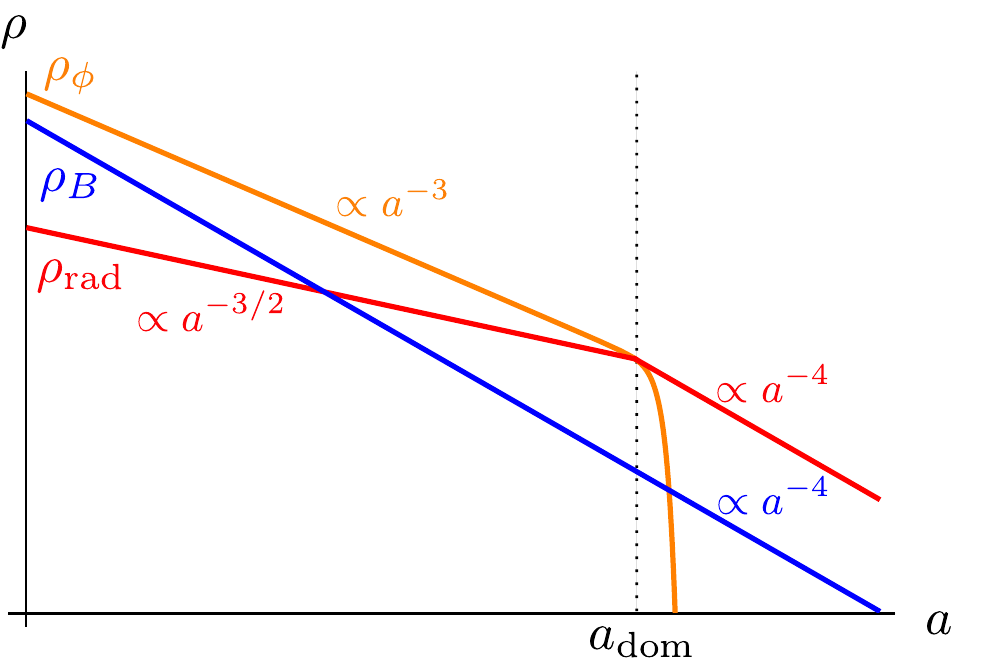}
  \end{center}
  \caption{Schematic of the evolution of energy densities in 
   the reheating and radiation-dominated epochs as functions of the 
   scale factor, in log-log scale. 
   Shown are the energy densities of an oscillating inflaton 
   (orange), radiation (red), and primordial magnetic fields (blue). 
   Here the magnetic field energy density is extrapolated back to the
   left edge of the plot (which corresponds to some time during reheating),
   but there is actually a cutoff corresponding to the time when the
   magnetic fields are generated. See the text for details.}
  \label{fig:rho-a}
  \end{center}
\end{figure}

Our bound also sets an upper limit on the scale~$H_i$ of magnetic field
generation, for a given value of~$m g^{-3/2}$. 
Let us also comment on other bounds on $H_i$. 
Firstly, as we are assuming the magnetic field generation to conclude
between the end of inflation and matter-radiation equality, 
the Hubble scale should lie within the range
$10^{-37}\, \ro{GeV} \lesssim H_i \lesssim 10^{14}\, \ro{GeV}$,
where the upper bound is the observational limit 
on the inflation scale.
Secondly, we have assumed that the magnetic field only gives a
subdominant contribution to the total energy density of the universe.
The time evolution of the energy densities in the post-inflation era
is illustrated in Figure~\ref{fig:rho-a}.
Here the orange line denotes the energy density of an oscillating
inflaton~$\rho_{\phi}$, and the red line denotes the radiation energy
density~$\rho_{\ro{rad}}$ which is created by the decay of the inflaton.
Since the generation of primordial magnetic fields and reheating are
in general different processes, we discuss the energy density of the
magnetic field~$\rho_{B}$ separately from $\rho_{\ro{rad}}$, and denote
it by the blue line in the figure. 
By extrapolating the magnetic energy
density back in time as $\rho_B \propto a^{-4}$, 
it can overtake $\rho_\phi$ in the reheating
epoch and dominate the universe, 
which would signal that the cosmological expansion history was once 
significantly affected by the magnetic field. 
However the scaling~$\rho_B \propto a^{-4}$ is actually cut off
at the time~$t_i$, and we constrain this by requiring
that the magnetic energy density never dominated the universe.
By using (\ref{eq:coll-H}) and (\ref{eq:adom-Tdom}),
the magnetic energy fraction at~$t_i$ is written as
\begin{equation}
\frac{\rho_{B i}}{3 M_{\ro{Pl}}^2 H_i^2}
\sim 10^{-19} 
\left(\frac{B_0}{10^{-15}\, \ro{G}} \right)^{2}
\ro{max.} \left\{
1, \left( \frac{H_i}{H_{\ro{dom}}} \right)^{2/3}
\right\},
\end{equation}
which is smaller than unity if $H_i \leq H_{\ro{dom}}$ and $B_0 \lesssim
10^{-6}\, \ro{G}$. However in cases with $H_i > H_{\ro{dom}}$,
then $\rho_{B i} < 3 M_{\ro{Pl}}^2 H_i^2$ requires\footnote{If
the coherence length of the primordial magnetic field happens to be
close to the CMB scales, the magnetic energy density is further
restricted from discussions on curvature perturbations.}
\begin{equation}
H_i \lesssim 10^{10}\, \ro{GeV}
\left(\frac{B_0}{10^{-15}\, \ro{G}} \right)^{-3}
\left(\frac{T_{\ro{dom}}}{1\, \ro{GeV}} \right)^{2},
\label{eq:maru-I}
\end{equation}
where we have rewritten $H_{\ro{dom}}$ in terms of $T_{\ro{dom}}$.
This condition is satisfied on the limits displayed in the plot;
e.g., on the red line ($T_{\ro{dom}} = 1 \, \ro{GeV}$), 
the condition is violated at $H_i \gtrsim 10^{10}\, \ro{GeV}$ which is
around the upper edge and beyond. 
Both upper limits on $H_i$, 
(\ref{eq:Hi-limit}) and (\ref{eq:maru-I}), 
are tightened by a larger value of~$B_0$;
we will see this explicitly below.

\subsection{Further Limits for Solitonic Monopoles}
\label{sec:f_limit}

For solitonic monopoles of spontaneously broken gauge theories, 
the mass limit~(\ref{eq:m-limit}) can be evaded 
by keeping the symmetry unbroken when the magnetic field is generated. 
However in such a case the monopoles produced later at the symmetry
breaking phase transition would induce a monopole problem,
unless the symmetry breaking scale is very low\footnote{A very low scale
post-inflation symmetry breaking might avoid cosmological issues while
allowing for an initially very strong primordial magnetic field to
survive until today, but we do not pursue this direction further herein.}
(see also discussions in Section~\ref{sec:remarks} and
Appendix~\ref{app:PT}).
Thus we can combine the requirement to avoid a post-inflation symmetry
breaking with the mass limit, and give further constraints for
solitonic monopoles. 

In the following, for concreteness, 
we study the vanilla 't Hooft--Polyakov monopole
of an SO(3) gauge theory spontaneously broken to
U(1)~\cite{tHooft:1974kcl,Polyakov:1974ek}. 
In this case the monopole mass is related to the vacuum expectation
value of a triplet Higgs field~$\sigma$, which we also refer to as the
symmetry breaking scale, via $m \sim g \sigma$ (the exact value depends
also on the Higgs self-coupling~\cite{Kirkman:1981ck}).
The magnetic charge is $g = 4 \pi / e$ in terms of the gauge
coupling~$e$.

\begin{figure}[htbp]
\centering
\subfigure[$T_{\ro{dom}} = 1\, \ro{GeV}$]{%
  \includegraphics[width=.4\linewidth]{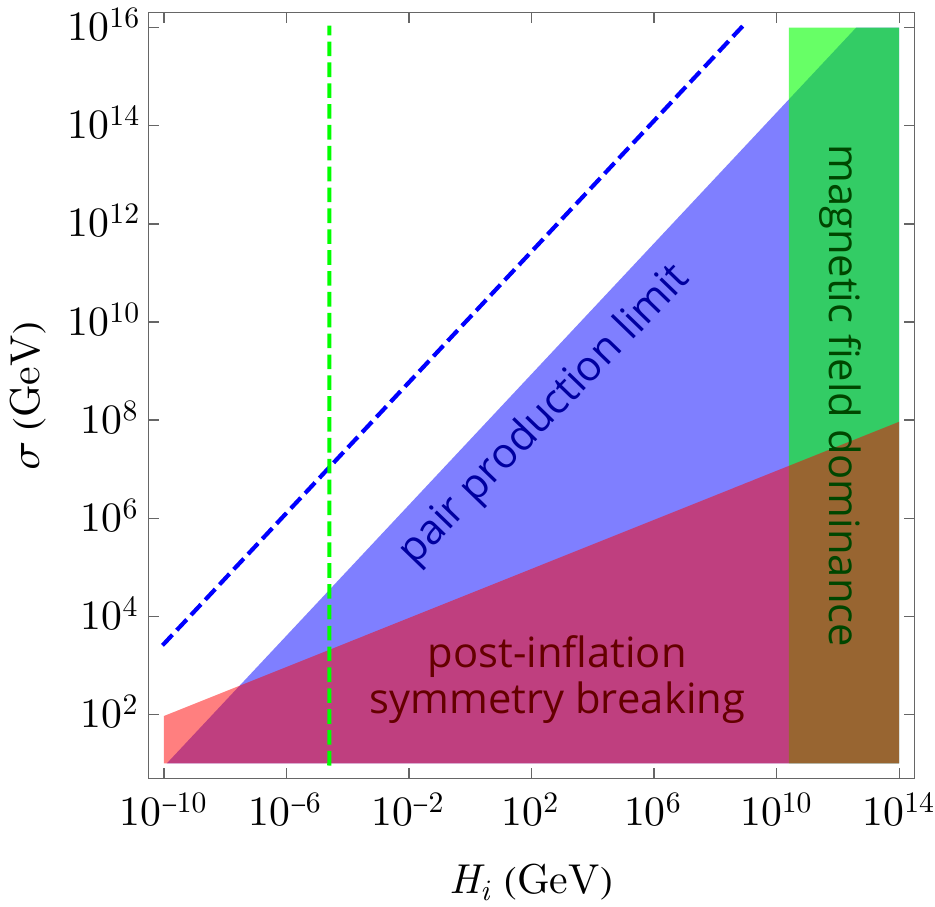}
  \label{fig:E0}}
\quad
\subfigure[$T_{\ro{dom}} = 10^4\, \ro{GeV}$]{%
  \includegraphics[width=.4\linewidth]{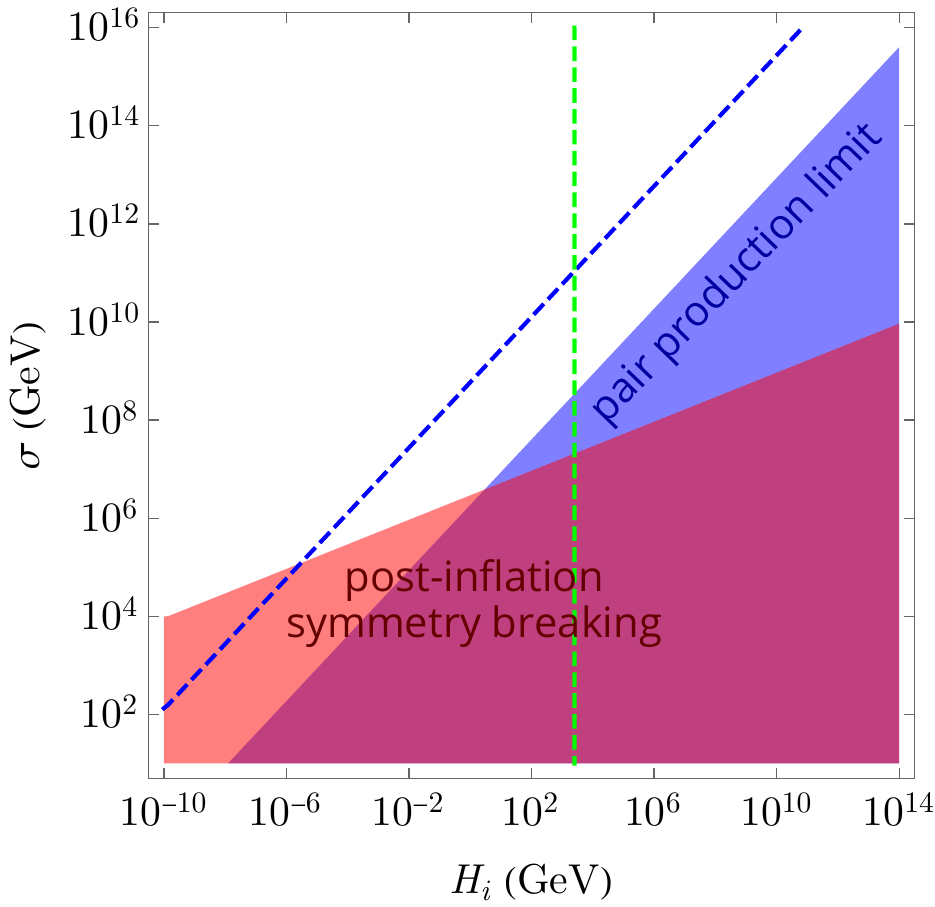}
  \label{fig:E4}}
\subfigure[$T_{\ro{dom}} = 10^8\, \ro{GeV}$]{%
  \includegraphics[width=.4\linewidth]{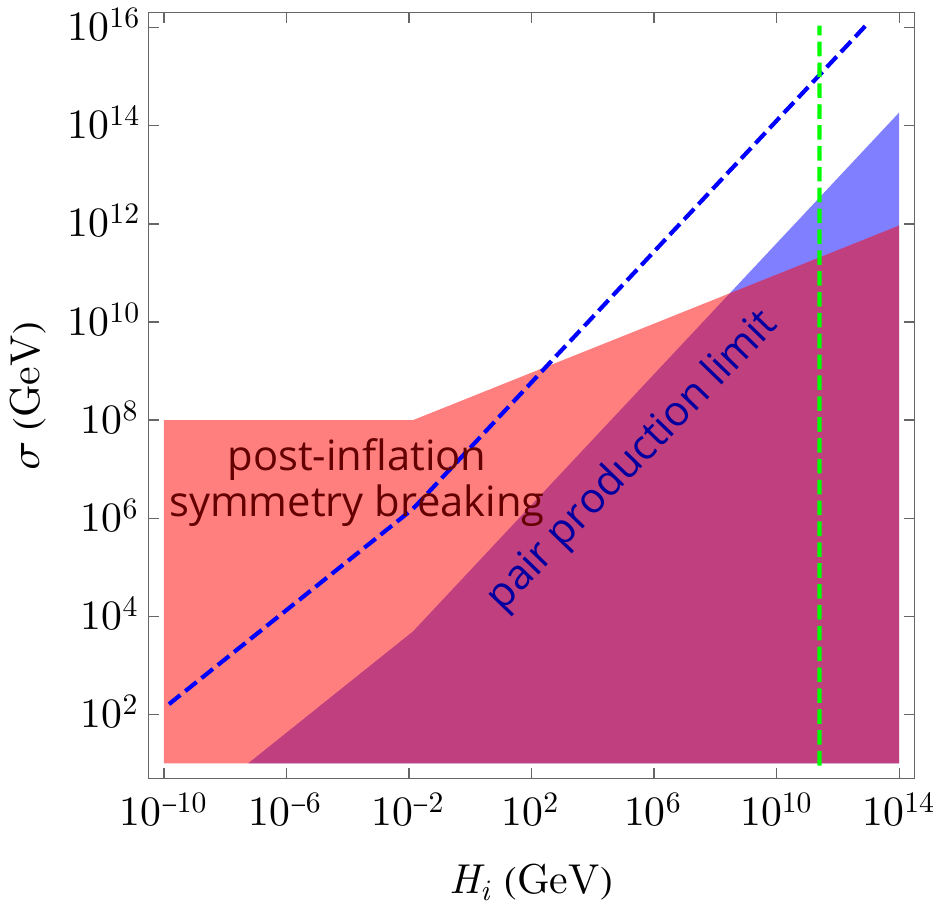}
  \label{fig:E8}}
\quad
\subfigure[$T_{\ro{dom}} = 10^{12}\, \ro{GeV}$]{%
  \includegraphics[width=.4\linewidth]{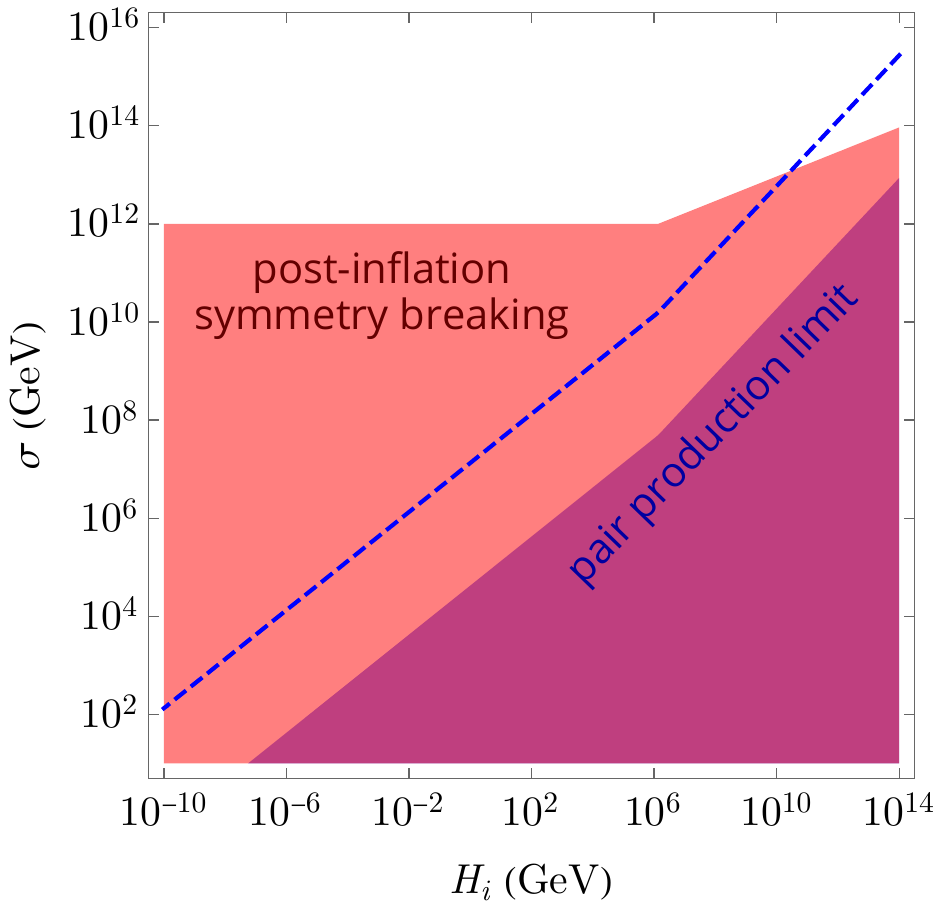}
  \label{fig:E12}}
 \caption{Parameter space of 't Hooft--Polyakov monopoles in the plane
 of the Hubble scale~$H_i$ 
 when primordial magnetic fields are initially generated,
 and the symmetry breaking scale~$\sigma$.
 The magnetic charge is fixed to $g = 10$, and the present-day magnetic
 field strength to $B_0 = 10^{-15}\, \ro{G}$.
 The cosmic temperature~$T_{\ro{dom}}$ when radiation domination
 begins is varied in the four plots. 
 The blue region violates the monopole mass limit~(\ref{eq:m-limit})
 obtained by the analyses of pair production in primordial magnetic fields.
 The green region is excluded by magnetic field dominance in the early
 universe.
 In the red region
 the symmetry breaking phase transition happens after inflation.
 For a stronger magnetic field $B_0 = 10^{-10}\, \ro{G}$, the left edges
 of the blue and green regions shift to the positions depicted by the
 dashed lines.}
\label{fig:sigma-Hi}
\end{figure}

To avoid a symmetry breaking after inflation, the symmetry breaking scale 
should be high enough to satisfy $\sigma > \ro{max.} \{ T,  H \}$ 
throughout the post-inflation universe.
During radiation domination, this implies
$\rho_B \sim B^2 \lesssim \rho_{\ro{rad}} \sim T^4 \lesssim \sigma^4 $
(here we neglect numerical coefficients).
Hence with $m \sim g \sigma$, we get $B \lesssim g B_{\star}$.
Thus one sees that if, say, $g = \mathcal{O} (10)$, 
then unless the inequalities are close to being saturated, 
the magnetic field in the symmetry broken phase is 
well below the threshold value~$B_\star$ for significant monopole
production (although there can still be non-negligible effects
below~$B_\star$ as discussed in the previous sections).
However this is no longer the case in the epoch prior to
radiation domination, where $\rho_B$ can be larger
than~$\rho_{\ro{rad}}$, while being smaller than the dominant
inflaton energy density (cf. Figure~\ref{fig:rho-a}).
This kind of situation can arise, for instance, in 
magnetic field generating mechanisms that invoke a 
violation of the Weyl invariance of the Yang--Mills
action (see e.g. \cite{Turner:1987bw,Ratra:1991bn});
these take place only in a cold universe such as during inflation, 
since otherwise electrically charged particles in the thermal plasma
freeze in the magnetic flux.
While such mechanisms are in operation, the energy density of the
magnetic field is typically much larger than that of the radiation
component.

In Figure~\ref{fig:sigma-Hi} we show the parameter space 
of 't Hooft--Polyakov monopoles in the $H_i$-$\sigma$ plane, 
where we took $m = g \sigma$, $g = 10$, and $B_0 = 10^{-15}\, \ro{G}$.
$T_{\ro{dom}}$ is varied in the four plots as 
$1\, \ro{GeV}$, $10^{4}\, \ro{GeV}$, $10^{8}\, \ro{GeV}$, and $10^{12}\,
\ro{GeV}$.
The blue region is excluded by the lower limit~(\ref{eq:m-limit}) on
the monopole mass, and corresponds to that shown in
Figure~\ref{fig:mg-Hi}.  
The green region violates the magnetic field energy
bound~(\ref{eq:maru-I}), which is seen only in the plot for $T_{\ro{dom}} =
1\, \ro{GeV}$, since for $T_{\ro{dom}} \gtrsim 10^{2}\, \ro{GeV}$ 
the upper limit on~$H_i$ from this bound exceeds the highest possible
inflation scale.
The red region shows where $\sigma < \ro{max.} \{T_i, T_{\ro{dom}} \}$,
indicating that the symmetry breaking takes place after inflation
and thus possibly gives rise to a monopole problem.
Here, $T_i$ is given in terms of $H_i$ through (\ref{eq:coll-H}) and
(\ref{eq:coll-T}).\footnote{A perturbative reheating is assumed here. If
instead the inflaton decays non-perturbatively (so-called
preheating), the evolution of the cosmic temperature at $T >
T_{\ro{dom}}$ could be modified.}
We do not show where $\sigma < \ro{max.} \{H_i, H_{\ro{dom}} \}$ since
it only gives constraints weaker than the other conditions in the
displayed parameter regions.


As we have already discussed, the requirement of 
$\sigma > \ro{max.} \{T_i, T_{\ro{dom}} \}$ gives a stronger constraint
than the mass limit~(\ref{eq:m-limit})
during radiation domination ($H_i < H_{\ro{dom}}$), and serves as the
dominant constraint in the entire displayed space in the plot for
$T_{\ro{dom}} = 10^{12}\, \ro{GeV}$. 
For the other plots with lower~$T_{\ro{dom}}$, the mass limit dominates
at $H_i \gg H_{\ro{dom}}$, i.e., if the primordial magnetic field is
generated long before radiation domination. 
The combination of $\sigma > \ro{max.} \{T_i, T_{\ro{dom}} \}$  
and the mass limit (\ref{eq:m-limit}) put severe constraints on symmetry
breaking at intermediate and low scales.
For instance, the necessary condition for $\sigma = 10^{8}\, \ro{GeV}$
to evade the two constraints is that $T_{\ro{dom}} < 10^8\, \ro{GeV}$
and $H_i \lesssim 10^3\, \ro{GeV}$ are both satisfied. 

The condition $\sigma > \ro{max.} \{T_i, T_{\ro{dom}} \}$ 
is independent of~$B_0$, 
while the limits (\ref{eq:m-limit}) and (\ref{eq:maru-I}) become stronger
for a larger~$B_0$.
In the plots we also show (\ref{eq:m-limit}) and
(\ref{eq:maru-I}) for $B_0 = 10^{-10}\, \ro{G}$, 
by the blue and green dashed lines, respectively.
With this larger~$B_0$, the monopole mass limit is tightened by about
two orders of magnitude, and overtakes the constraint from 
$\sigma > \ro{max.} \{T_i, T_{\ro{dom}} \}$ in a wider parameter range. 
The magnetic energy bound is also tightened, and is seen to constrain
the high-$H_i$ regions in the plots with 
$T_{\ro{dom}} $ up to $ 10^8\, \ro{GeV}$.  

Note that, to keep the discussion general, we have not specified the
inflation scale. We focused on the time $t_i$ at the end of magnetic
field generation which may coincide with the end of inflation, but can
also be at some later time. 
Accordingly, we only imposed 
$\sigma > \ro{max.} \{T_i, H_i, T_{\ro{dom}}, H_{\ro{dom}} \}$,
i.e. the symmetry to be broken by the time when magnetic field generation
completes or radiation domination begins, whichever happens earlier,
instead of imposing $\sigma > \ro{max.} \{ T, H\}$ since the end of
inflation. 
The actual lower bound on $\sigma$ for evading a post-inflation
symmetry breaking would be tighter than shown in the figure, 
if radiation domination and magnetic field generation take place long
after the end of inflation.

\section{Conclusions}
\label{sec:conc}

We showed that the process of pair production in primordial magnetic fields
provides an excellent opportunity to confront magnetic monopoles with
astrophysical observations. 
We analyzed two major consequences of the monopole pair production:
(i) Primordial magnetic fields dissipate energy by producing the
monopole pairs and subsequently accelerating them.
This fact that the field self-screens yields a
consistency condition for primordial magnetic fields to survive until
today and explain the observed magnetic fields.
(ii) The pair produced monopoles can give rise to a new type of 
monopole problem, which gives a cosmological bound on
monopoles and primordial magnetic fields.
After evaluating the constraints from each effect,
we used the most conservative bound on the
primordial magnetic field amplitude~(\ref{eq:smaller-than-B_up}) to
derive a lower limit on the monopole mass~(\ref{eq:m-limit}):
\begin{equation}
m \gtrsim 10^{13}\, \ro{GeV}
\left( \frac{g}{20}  \right)^{3/2}
\left( \frac{B_0}{10^{-15}\, \ro{G}} \right)^{1/2}
\left( \frac{H_i}{10^{14}\, \ro{GeV}} \right)^{1/2}
\ro{max.} \left\{
1, \left( \frac{H_i}{H_{\ro{dom}}} \right)^{1/6}
\right\}.
\end{equation}
Here $g$ is the magnetic charge of the monopole, $B_0$ is the
present-day magnetic field strength, 
$H_i$ is the Hubble scale when the primordial magnetic field is
initially generated, 
and $H_{\ro{dom}}$ is the Hubble scale when radiation domination begins.
This limit also serves as an upper bound on the scale of magnetic field
generation.
A primordial magnetic field that seeds the observationally suggested 
intergalactic magnetic fields of $B_0 \gtrsim 10^{15}\, \ro{GeV}$,
imposes constraints on monopoles for a wide mass range 
(Figure~\ref{fig:mg-Hi}).
This also sets a constraint on grand unified theories, which is
particularly severe for models with intermediate and low scale symmetry
breaking. 
Moreover, we showed that even superheavy monopoles of 
$m \sim 10^{16}\, \ro{GeV}$ can be abundantly produced if primordial
magnetic fields exist at sufficiently high redshifts. 

It is also important to know the exact abundance of monopoles produced in
primordial magnetic fields, in order to make concrete predictions for
monopole search experiments.
Because the pair production rate depends exponentially on the magnetic
field, the threshold field strengths for magnetic self-screening and
monopole overabundance are typically of the same order. 
Moreover, this threshold value may only marginally satisfy the weak field
condition invoked in the instanton calculation of the pair production
rate.   
Therefore a precise evaluation of the monopole abundance
would require solving the full system including the
backreaction from the monopoles on the magnetic field, with possible
corrections to the pair production rate for marginally weak fields.
Additional effects that deserve careful studies are listed in
Section~\ref{sec:summary}. 
Taking into account all of them can be non-trivial, however
there may be fortunate circumstances where some effects 
decouple from the rest to simplify the analysis.
Alternatively, if some effects can be argued
to work only in a certain direction, such as to always 
enhance the pair production, then one can ignore those effects
to derive conservative limits,
which is the strategy adopted in this paper.
Despite being conservative, our constraint should be useful since it
applies to monopoles with a  wide mass range, including superheavy ones
that are practically impossible to probe in colliders. 

Magnetic field generation in the early universe can be accompanied
by a simultaneous generation of electric fields, which are considered to
eventually short out during the reheating process.
It would be interesting to study monopole production in
primordial magnetic and electric fields before the latter vanish
(see e.g. \cite{Kim:2003qp,Tanji:2008ku} for studies of Schwinger pair
production in electric and magnetic fields).
We also note that our analyses can be extended to the production
of dyons~\cite{Schwinger:1969ib} from primordial electromagnetic fields.
Finally we note that the pair production in primordial fields
works equally effective for monopoles and magnetic fields of
hidden U(1) gauge fields, therefore it can provide a new production
mechanism for hidden monopole dark matter~\cite{Murayama:2009nj}.

Ultimately, one wishes to probe theories of monopoles and quantum vacuum
instability via astrophysical measurements of cosmological magnetic
fields, and in  turn, to reveal the origin of cosmological magnetic
fields by studying monopole pair production. 
This paper serves as a first step towards this goal.

\section*{Acknowledgments}

I am grateful to Nobuhiro Maekawa for comments on the manuscript, and to
Masaki Shigemori and Hiroyuki Tashiro for helpful discussions. 
This research is supported by
Grant-in-Aid for Scientific Research on Innovative Areas No.~16H06492.

\appendix

\section{General Discussion of Magnetic Field Dissipation by Monopoles}
\label{app:E}

We present a general discussion on the dissipation of cosmological
magnetic fields by monopoles.

\subsection{General Formalism}

The physical energy density of a spatially homogeneous magnetic field
in an FRW background universe obeys
\begin{equation}
 d\left[ \rho_B(t) a(t)^3 \right]
= -P_B(t) \, d \left[a(t)^3 \right]
- 2 m \Gamma(t) a(t)^3 dt
- 2 g B(t) dt \int^t_{-\infty} dt' \, a(t')^3  \Gamma (t') v(t',t),
\label{eq:A.1}
\end{equation}
where $P_B$ is the pressure of the magnetic fluid, and
$\Gamma$ is the rate of monopole-antimonopole pair production by the
magnetic field. 
The second term in the right hand side denotes the magnetic field energy
being depleted by~$2m$ for the production of each pair, assuming the
pairs to be produced at rest (this term corresponds
to~(\ref{eq:Gamma_m})). 
The third term represents the energy loss by accelerating the population
of pairs produced from the infinite past to time~$t$, 
where it should be noted that
$a(t')^3 \Gamma (t') dt'$ gives the comoving number density of pairs
produced between $t'$ and $t' + dt'$.
Moreover, we used $v(t', t)$ to denote the velocity of monopoles produced
at~$t'$, measured at~$t$ ($\geq t'$), 
in the direction of the magnetic field
(antimonopoles are taken to have charge $-g$ and velocity $-v (t',t)$).
We have ignored monopole-antimonopole annihilation.

Using $\rho_B = B^2 / 2$, 
and supposing a barotropic equation of state
$P_B/\rho_B = (2p/3) - 1$ for the magnetic fluid, 
which amounts to supposing $B \propto a^{-p}$ in the absence of
monopole production, 
then (\ref{eq:A.1}) is rewritten as
\begin{equation}
 \frac{\dot{\rho}_B}{\rho_B}
=  -\Pi_{\ro{red}} -\Pi_{\ro{prod}} -\Pi_{\ro{acc}}, 
\label{eq:gen-for}
\end{equation}
with the damping rates due to redshifting, monopole production, and
monopole acceleration:
\begin{equation}
\Pi_{\ro{red}}(t) = 2 p H(t),
\quad
\Pi_{\ro{prod}}(t) = \frac{4 m \Gamma (t) }{B(t)^2},
\quad
\Pi_{\ro{acc}}(t) = \frac{4 g}{a(t)^3 B(t)}
\int^t_{-\infty} dt' \, a(t')^3  \Gamma (t') v(t',t).
\label{eq:rates}
\end{equation}
The depletion of the magnetic field energy can be studied by solving
this equation, combined with the expression~(\ref{eq:exact_v}) for
$v(t',t)$ given below.

\subsection{Monopole Velocity}
\label{app:velocity}

The motion of a monopole with magnetic charge~$g$ in a homogeneous
magnetic field and an FRW background spacetime is described by the
equation of motion,
\begin{equation}
\frac{m}{a} \frac{d}{dt} 
\left( a \gamma v \right) =  g B  ,
\quad
\gamma = \frac{1}{\sqrt{1-v^2}} ,
\end{equation}
where $v$ is the velocity in the direction of the magnetic field, 
and we neglected motion perpendicular to the magnetic field.
This is integrated as
\begin{equation}
 \gamma (t', t) \, v (t', t) = 
 \frac{g}{m a(t)} \int_{t'}^{t} dt'' \, a(t'') B(t'') ,
\end{equation}
or equivalently,
\begin{equation}
v(t', t) = 
\frac{
\frac{g}{m a(t)}
\int_{t'}^{t}
dt'' \, a(t'') B(t'')
}{
\sqrt{1 + \left[ \frac{g}{m a(t)}
\int_{t'}^{t}
dt'' \, a(t'') B(t'')
\right]^2}
}.
\label{eq:exact_v}
\end{equation}
Here $t$ is the time when the velocity is measured, and 
$t'$ is when the monopole was initially produced at rest.

For example, if the magnetic field scales as $B \propto a^{-p}$, and
the Hubble rate as $H \propto a^{-3 (1+w)/2}$ with a constant
equation of state~$w$, 
the integral can be directly performed as
\begin{equation}
\gamma v =
\frac{1}{1-\nu } \frac{g B'}{m H'}
\left[
\left(\frac{a'}{a}\right)^\nu  - \left(\frac{a'}{a}\right)
\right],
\quad
\nu = p - \frac{3 (1+w)}{2},
\label{eq:gv-w}
\end{equation}
where $a = a(t)$, $a' = a(t')$, etc.
The behavior of $\gamma v$ in the asymptotic future is as follows:
If $\nu > 0$, it decays in time as either
$\gamma v \propto a^{-1}$ ($\nu > 1$), 
$\gamma v \propto a^{-1} \ln a$ ($\nu = 1$), 
or $\gamma v \propto a^{-\nu}$ ($0 < \nu < 1 $).
If $\nu < 0$, it grows as $\gamma v \propto a^{\abs{\nu}}$.
If $\nu = 0$, i.e. the equation of state of the universe
equals that of the magnetic fluid, then it asymptotes to a constant value
$\gamma v \rightarrow g B' / m H'$.

\subsection{Case Studies of Dissipation by Monopole Acceleration}

Let us evaluate the dissipation rate by monopole
acceleration, $\Pi_{\ro{acc}}$, under the simplifying 
assumption that the magnetic field is suddenly switched on at
time~$t_i$, then subsequently redshifts as $B \propto a^{-p}$ with a
positive~$p$ of order unity (while the dissipation by monopoles is
negligible).
Then with $\Gamma$ of the form~(\ref{eq:Gamma}), the integral
for~$\Pi_{\ro{acc}}$ is dominated by the contribution from 
$t_i \leq t' \lesssim t_i + \Delta t_{\Gamma i}$
where $\Delta t_{\Gamma i} = \epsilon_i / p H_i$
($\ll 1 / H_i$, see discussions below~(\ref{eq:epsilon})).
Ignoring the variation of $a^3 \Gamma$ during this period
(which implicitly assumes 
$\Delta t_{\Gamma i} < \Pi_{\ro{prod}\, i}^{-1}, \Pi_{\ro{acc}\,
i}^{-1}$), we get 
\begin{equation}
 \Pi_{\ro{acc}}(t) \simeq \frac{4 g \Gamma_i }{B(t)}
\left( \frac{a_i}{a(t)}\right)^3
\int^{t_i + \Delta t_{\Gamma i}}_{t_i} dt' v(t',t),
\label{eq:A.8}
\end{equation}
for $t \geq t_i + \Delta t_{\Gamma i}$.
For a further evaluation, we consider some limiting cases below.

\subsubsection{Monopoles Produced on the Spot}
\label{app:spot}

We start by considering the times 
$\Delta t_{\Gamma i} \ll t - t_i \lesssim 1 / H_i$,
which are within a Hubble time after the magnetic field is switched on. 
During this period the expansion of the universe can be ignored.
Further supposing $B$ to be nearly constant (which amounts to ignoring
the backreaction from the monopoles), 
then (\ref{eq:A.8}) is approximated as
\begin{equation}
 \Pi_{\ro{acc}}(t) \simeq \frac{4 g \Gamma_i }{B_i}
\int^{t_i + \Delta t_{\Gamma i}}_{t_i} dt' v(t',t).
\end{equation}
Likewise, the monopole velocity (\ref{eq:exact_v}) in the integral is
approximated as 
\begin{equation}
 v (t', t) \simeq \frac{\frac{gB_i}{m} (t-t')}{\sqrt{1 + 
\left[ \frac{gB_i}{m} (t-t') \right]^2 }}.
\end{equation}

If the initial magnetic field is sufficiently strong such that the 
monopoles are relativistic at the time of consideration, 
i.e. $(gB_i / m) (t-t_i) \gg 1$, then 
$\int dt' v \simeq \Delta t_{\Gamma i}$
and we obtain
\begin{equation}
 \Pi_{\ro{acc}} \simeq
\frac{4 g^2 \Gamma_i}{\pi p m^2 H_i}.
\end{equation}
This matches with the dissipation rate 
$\abs{(\dot{\rho}_B)_{\ro{R}} / \rho_B}$ of
(\ref{eq:Gamma_dis}) with the substitution of (\ref{eq:n-at-the-time}) 
and $t = t_i$.

On the other hand for non-relativistic monopoles, 
i.e. $(gB_i / m) (t-t_i) \ll 1$, 
then $\int dt' v \simeq (gB_i / m) \Delta t_{\Gamma i} (t-t_i)$
and we get
\begin{equation}
 \Pi_{\ro{acc}} \simeq \frac{4 g^3 B_i \Gamma_i}{ \pi p m^3
  H_i}  (t-t_i) .
\end{equation}
Using this to solve 
$\dot{\rho}_B / \rho_B = - \Pi_{\ro{acc}}$, one obtains
$\rho_B = \rho_{B i} \exp [ -\{ (t-t_i) / \Delta t_{\ro{acc}} \}^2 ]$
with the dissipation time scale given by
\begin{equation}
 \Delta t_{\ro{acc}} = \left(
\frac{\pi p m^3 H_i}{2 g^3 B_i \Gamma_i}
\right)^{1/2}.
\end{equation}
This matches with 
$ \Delta t_{\ro{NR}}$ derived in 
(\ref{eq:t_dis}) with the substitution of (\ref{eq:n-at-the-time})
and $t = t_i$.

\subsubsection{Monopoles Produced in the Past}
\label{app:preexist}

We now consider the times $ t - t_i \gtrsim 1 / H_i$.
Here we assume for simplicity that
$v(t', t) \simeq v(t_i, t)$
for $ t_i \leq t' \leq t_i + \Delta t_{\Gamma i}$,
i.e., most monopoles have the same velocity.
Then $\int dt' v \simeq \Delta t_{\Gamma i} v(t_i, t)$, which gives
\begin{equation}
 \Pi_{\ro{acc}} \simeq
\frac{4 g^2 \Gamma_i}{\pi p m^2 H_i} \frac{B_i}{B(t)}
\left(\frac{a_i}{a(t)}\right)^3 v(t_i, t).
\end{equation}
This denotes the rate of magnetic field dissipation by
accelerating monopoles that have been produced at around the initial
time~$t_i$, which is well separated from the time~$t$ of consideration.
We did not discuss this effect in the main text.

Let us focus on the time evolution of 
$\Pi_{\ro{acc}}$ with respect to the redshifting
rate~$\Pi_{\ro{red}}$,
\begin{equation}
 \frac{\Pi_{\ro{acc}}}{\Pi_{\ro{red}}} \propto 
\frac{v(t_i, t)}{B(t) a(t)^3 H(t)}. 
\end{equation}
For $v$, we can use (\ref{eq:gv-w}) 
when the universe has an equation of state~$w$,
and while the magnetic field scales as $B \propto a^{-p}$.
For instance, with $ p = 2$ and $w = 1/3$,
the velocity~$v$ approaches a constant value and thus the ratio grows
asymptotically as
$ \Pi_{\ro{acc}} / \Pi_{\ro{red}} \propto a $.
In this case, even if the dissipation by monopole acceleration is
initially negligible, 
it can become important at later times. 
For $p = 2$ and $w = 0$, we get
$ \Pi_{\ro{acc}} / \Pi_{\ro{red}} \propto a^{1/2} $ 
while the monopoles are relativistic ($v \simeq 1$), and 
$ \Pi_{\ro{acc}} / \Pi_{\ro{red}} = \ro{const.} $ 
when non-relativistic ($v \propto a^{-1/2}$).
It would be interesting to perform a systematic study of the dissipation
effect by monopoles produced in the past.

\section{Hubble Scale During and After Reheating}
\label{app:reheating}

We give the expressions for the Hubble scale and cosmic temperature
as functions of redshift during the reheating epoch and the subsequent 
radiation-dominated epoch.
Here we assume that after inflation ends~(at $t_{\ro{end}}$), 
the universe is initially dominated by an oscillating inflaton field,
which undergoes perturbative decay into radiation;
the radiation component eventually comes to dominate the
universe~($t_{\ro{dom}}$), until it gives way to matter domination at 
matter-radiation equality~($t_{\ro{eq}}$). 
A case of an instantaneous reheating, i.e., a sudden decay of the
inflaton at the end of inflation, is handled by setting 
$t_{\ro{end}} = t_{\ro{dom}}$ in the following discussions.

During radiation domination ($t_{\ro{dom}} \ll t \ll t_{\ro{eq}}$),
we have $3 M_{\ro{Pl}}^2 H^2 \simeq \rho_{\ro{rad}}$
where $\rho_{\ro{rad}} = (\pi^2/30) g_* T^4$ is the radiation energy
density and $T$ is the radiation temperature.
Combining this with the assumption that the entropy is conserved until
today, namely, that the entropy density redshifts as
$s = (2 \pi^2/45) g_{*s} T^3 \propto a^{-3}$, one obtains
\begin{equation}
  H \simeq \left( \frac{45}{128 \pi^2} \right)^{1/6}
\frac{g_*^{1/2}}{g_{*s}^{2/3}} \frac{s_0^{2/3}}{M_{\ro{Pl}}}
\left( \frac{a_0}{a} \right)^2,
\quad
 T \simeq \left( \frac{45}{2 \pi^2} 
\frac{s_0}{g_{*s}} 
\right)^{1/3}
\frac{a_0}{a}.
\label{eq:HT_RD}
\end{equation}
The subscript~``$0$'' denotes quantities in the present universe.

When the universe is dominated by an oscillating inflaton field
($t_{\ro{end}} \ll t \ll t_{\ro{dom}}$),
it is effectively matter-dominated and thus $H \propto a^{-3/2}$. 
The radiation density during this epoch is sourced by the perturbative
decay of the inflaton, and thus redshifts as 
$\rho_{\ro{rad}} \propto a^{-3/2}$
when ignoring the time dependence of~$g_*$;
this can be checked by solving the continuity equation
$\dot{\rho}_{\ro{rad}} + 4 H \rho_{\ro{rad}} = \Gamma_\phi \rho_\phi$
with $\Gamma_{\phi}$ being the inflaton decay rate,
and the energy density of the inflaton 
$\rho_\phi = \rho_{\phi \ro{end}} (a_{\ro{end}} / a)^3 
e^{-\Gamma_\phi (t - t_{\ro{end}})}$~\cite{Kolb:1990vq}.
Hence the radiation temperature redshifts as $T \propto a^{-3/8}$.

Connecting the scaling behaviors in the two epochs at~$t_{\ro{dom}}$,
the Hubble rate and radiation temperature
during the radiation-dominated epoch ($t_{\ro{dom}} < t < t_{\ro{eq}}$) 
and reheating epoch ($t_{\ro{end}} < t < t_{\ro{dom}}$) 
are collectively written as:
\begin{gather}
 H \sim H_{\ro{dom}} \, 
\ro{min.}\left\{
\left( \frac{a_{\ro{dom}}}{a} \right)^{2},
\left( \frac{a_{\ro{dom}}}{a} \right)^{3/2}
\right\},
\label{eq:coll-H}
\\
 T \sim T_{\ro{dom}} \,
\ro{min.}\left\{
\left( \frac{a_{\ro{dom}}}{a} \right),
\left( \frac{a_{\ro{dom}}}{a} \right)^{3/8}
\right\},
\label{eq:coll-T}
\end{gather}
where we have ignored the time variation of $g_{*(s)}$ in (\ref{eq:HT_RD}).
The relations between $H_{\ro{dom}}$, $T_{\ro{dom}}$, and $a_{\ro{dom}}$
can be obtained by extrapolating (\ref{eq:HT_RD}) to the
time~$t_{\ro{dom}}$; after plugging in numbers for $M_{\ro{Pl}}$
and the cosmological parameters one gets
\begin{equation}
 \frac{a_0}{a_{\ro{dom}}} \sim 10^{29} 
\left(\frac{H_{\ro{dom}}}{10^{14}\, \ro{GeV}} \right)^{1/2} ,
\quad
T_{\ro{dom}} \sim 10^{16}\, \ro{GeV}
\left(\frac{H_{\ro{dom}}}{10^{14}\, \ro{GeV}} \right)^{1/2}.
\label{eq:adom-Tdom}
\end{equation}
We remark that these depend only weakly on~$g_{*(s)}$, hence its
detailed value does not affect the order-of-magnitude estimates.

\section{Monopole Abundance Produced at Phase Transitions}
\label{app:PT}

In this appendix we consider solitonic monopoles produced at a symmetry
breaking phase transition that happens after inflation.
Hence the critical temperature~$T_{\ro{c}}$ at the phase
transition is assumed to be
lower than the maximum temperature achieved during
reheating, or the inflationary Hubble scale, i.e.,
$T_{\ro{c}} < \ro{max.}\{ T_{\ro{max}}, H_{\ro{inf}} \}$.
We consider a post-inflation history as discussed in
Appendix~\ref{app:reheating}, and obtain lower limits on the 
monopole abundance for cases where the phase transition takes place during
the reheating epoch, and during the radiation-dominated epoch.

Considering that at least one monopole or antimonopole is created within
a Hubble volume after the phase transition, the monopole number
density~$n_M$ follows $ n_{M \ro{c}} \geq H_{\ro{c}}^3$,
where the subscript~``c'' denotes quantities at the phase transition.
(Here we only compute a lower bound, but the actual density 
can be computed by evaluating the correlation length as discussed
in~\cite{Kibble:1976sj,Zurek:1985qw,Murayama:2009nj}.)
Then supposing that monopole-antimonopole annihilation is negligible,
and that the monopoles today are non-relativistic, 
the lower bound on the relic density is
\begin{equation}
 \rho_{M 0} = m \, n_{M0} \geq
 m H_{\ro{c}}^3 \left( \frac{a_{\ro{c}}}{a_0} \right)^3.
\end{equation}

If the phase transition happens during the radiation-dominated epoch, 
$t_{\ro{dom}} < t_{\ro{c}} < t_{\ro{eq}}$,
by rewriting the Hubble rate and redshift in terms of the cosmic
temperature using (\ref{eq:coll-H}), (\ref{eq:coll-T}), and
(\ref{eq:adom-Tdom}), one finds for the relic abundance,
\begin{equation}
 \Omega_M h^2 \gtrsim 10^{-1} 
\left( \frac{m}{10^{13}\, \ro{GeV}} \right)
\left( \frac{T_{\ro{c}}}{10^{11}\, \ro{GeV}} \right)^3.
\label{eq:MAPT-after-tdom}
\end{equation}

On the other hand if the phase transition happens prior to radiation
domination, $t_{\ro{end}} < t_{\ro{c}} < t_{\ro{dom}}$,
\begin{equation}
 \Omega_M h^2 \gtrsim 10^{-1}
\left( \frac{m}{10^{13}\, \ro{GeV}} \right)
\left( \frac{T_{\ro{c}}}{10^{11}\, \ro{GeV}} \right)^3
\left( \frac{T_{\ro{c}}}{T_{\ro{dom}}} \right).
\label{eq:MAPT-before-tdom}
\end{equation}
Compared to~(\ref{eq:MAPT-after-tdom}), this lower bound is enhanced by
$T_{\ro{c}} / T_{\ro{dom}}$.
This can be understood from the fact that for the same critical
temperature~$T_{\ro{c}}$, the Hubble scale at the phase
transition~$H_{\ro{c}}$ is larger during inflaton domination
than during radiation domination, 
and thus the number of monopoles is enhanced.

\bibliographystyle{JHEP}
\bibliography{monopole}

\end{document}